%% file: journal.tex
\documentclass[draftcls,12pt,onecolumn]{IEEEtran}  

\usepackage{graphicx}  
\usepackage{psfrag}    
\usepackage{subfigure} 
\usepackage{amsmath}   
\usepackage{amssymb}   
\usepackage{eucal}     
\newtheorem{lemma}{Lemma}
\newtheorem{proposition}{Proposition}
\newtheorem{definition}{Definition}
\newtheorem{corollary}{Corollary}

\newtheorem{theorem}{Theorem}

\input{rv_defs}
\begin{document}

\title{Variations on Information Embedding in \\
Multiple Access and Broadcast Channels}


\author
{{Shivaprasad Kotagiri~\IEEEmembership{Student Member,~IEEE}, and 
\\ J. Nicholas Laneman,~\IEEEmembership{Senior Member,~IEEE}}
 \thanks{Manuscript received \today.}
 \thanks{This work has been supported in part by NSF 
 Career Grant and  the State of Indiana through
 the Twenty-First Century Research and Technology Fund.}
 \thanks{Parts of this work were presented at Allerton 2005
 and IEEE ISIT 2006.}
 \thanks{Shivaprasad Kotagiri and J. Nicholas 
 Laneman are with Department of Electrical Engineering,
 University of Notre Dame, Notre Dame, IN 46556, 
 Email: \texttt{\{skotagir, jnl\}@nd.edu}}}

\maketitle
\IEEEpeerreviewmaketitle

\begin{abstract}
Information embedding (IE) is the transmission of 
information within a host signal subject to a
distortion constraint. There are two types of 
embedding methods, namely irreversible IE and reversible IE, 
depending upon whether or not the host, as well as the message,
is recovered at the decoder. In irreversible IE, only 
the embedded message is recovered at the decoder, 
and in reversible IE, both the message and the host are recovered
at the decoder. This paper considers combinations 
of  irreversible and reversible IE  in 
multiple access channels (MAC) 
and physically degraded broadcast channels (BC).

This paper first considers MAC IE in which separate encoders embed  
their messages into their host
signals subject to distortion constraints.
The embedded signals from the two encoders are 
transmitted to a single decoder across a MAC.
This paper study the capacity region in three cases:
A) no host recovery at the decoder,
B) lossless recovery of one
host at the decoder, and
C) lossless recovery of both hosts at the      
decoder.  For the cases A and B, inner 
bounds on the respective capacity regions are developed. 
For the case C, inner and outer bounds on the capacity region are
developed and the capacity region is obtained 
if the hosts are independent.

This paper also considers BC IE  in which two messages intended for 
separate decoders are embedded into a given host sequence by a 
single encoder subject to a distortion constraint.
This paper study the capacity region for degraded BC in four cases: 
$A'$) lossless recovery of the host sequence at neither of the decoders,
$B'$) lossless recovery of the host sequence at only  the better decoder,
$C'$) lossless recovery of the host sequence at both decoders, and
$D'$) lossless recovery of the host sequence at only  the worse decoder.
For the cases $A'$ and $B'$, inner and outer bounds on the 
respective capacity regions are developed. For the cases $C'$ and $D'$, 
the respective capacity regions are obtained.
\end{abstract}


%
\IEEEpeerreviewmaketitle

\section{Introduction} \label{sec:intr}
Information embedding (IE) is the reliable 
transmission of information
within a host signal subject to a distortion constraint.
IE is a recent area of digital media 
research with many applications including
active and passive copyright protection
(digital watermarking); steganography;
embedding important control, descriptive reference 
information into a given signal;
digital upgrades of communication infrastructure;
and covert communications 
\cite{bchen00, chen01:it, anderson98:jsac,swanson98:picc}.
The main idea of IE is that the host signal
can carry different messages at the same time by
allowing a small amount of distortion that can be tolerated at
the intended receiver for the host signal.
It has been observed  
that IE is closely related to state-dependent channel models with 
state known non-causally at the encoder \cite{gelfand80:pcit, mcosta83:it} 
\cite{bchen00,chen01:it,moulin03:it}.


\subsection{Forms of IE}
In IE, a message $\rvw$ is embedded  into a host signal $\rvs^n$ 
such that the embedded signal $\rvx^n$ is close to $\rvs^n$
under some prescribed distortion measure $d(\cdot,\cdot)$, i.e.,
$\mathbb{E}d(\rvx^n,\rvs^n) \leq \Delta$.  The decoder receives $\rvy^n$, 
which is drawn according a probability law $p(\svy^n|\svx^n,\svs^n)$ 
for given $\rvx^n$ and $\rvs^n$. Throughout the paper, we focus on
the discrete memoryless case without feedback and denote the channel law
by $p(\svy|\svx,\svs).$ Based upon whether or not the decoder recovers 
the host signal in the sense of probability of error going to zero,  
there are two important types of IE, 
namely \textit{irreversible} and \textit{reversible} IE.

In irreversible IE, the decoder is only concerned with 
reliable decoding of the message embedded
in the host from the received sequence $\rvy^n$ 
\cite{bchen00, chen01:it, moulin03:it, acohen02:it}.
The irreversible IE capacity of a single-user model is given by
$$C(\Delta) = \max_{p(\svu,\svx|\svs):~\mathbb{E}d(\rvx,\rvs) \leq \Delta}
[\mathbb{I}(\rvu ;\rvy)-\mathbb{I}(\rvu ; \rvs)],$$
where $\rvu$ is an auxiliary random variable with $|\calU| \leq |\calX||\calS|$.
To achieve the capacity,
Gel'fand-Pinsker coding \cite{gelfand80:pcit} is used at the encoder such
that the distortion  between $\rvx^n$ and $\rvs^n$ satisfies 
the constraint $\Delta$.

In reversible IE, the decoder is concerned with 
lossless recovery of the host as well as reliable decoding of
the embedded message in the host from the received sequence $\rvy^n$ 
 \cite{kalker02:picdsp, kalker03:pspie}.
Reversible IE is useful for cases in which little or no degradation
of the host signal is allowed, with applications in
military and medical imagery, and multimedia archives 
of valuable original works.
The reversible IE capacity is given by
$$C(\Delta) = \max_{p(\svx|\svs):~\mathbb{E}d(\rvx,\rvs) \leq \Delta}
[\mathbb{I}(\rvx,\rvs ;\rvy)-\mathbb{H}(\rvs)].$$
To achieve the above capacity expression, superposition 
coding is used at the encoder such
that the distortion constraint is satisfied, 
i.e., $\mathbb{E}[d(\rvx,\rvs)] \leq \Delta $.

This paper focuses on IE in multi-user channels such as
multiple access channels (MAC) and broadcast channels (BC).
We focus on MAC IE with lossless recovery of
\textit{some} host sequences at the decoder and BC IE   
with lossless host recovery at \textit{some} decoders, but the techniques
can also be applied to other multi-user scenarios.
In single-user IE, substantial results have been developed,
but multi-user IE scenarios have not been as extensively studied.
Information theoretic study of single-user public and 
private watermarking systems is studied in
\cite{somekh-baruch03:it, somekh-baruch04:it, somekh-baruch05:it}.
Joint IE and lossy compression is studied in \cite{maor05:it, amaor05:it}
and joint watermarking and encryption is studied in \cite{merhav06:it}.
Multi-user models with state available at the encoders are studied
in \cite{gelfand83:pisit,kim04:pisit,khisti04:pisit},
\cite{kotagiri04:pallerton, ysteinberg05:it},
\cite{ycemal05:it, jafar06:it, kotagiri06:it},
\cite{baruch06:pallerton, kotagiri07:pisit, baruch07:pisit}.
As in single-user case, there is a close relationship between
multi-user models with non-causal state at the encoders and multi-user IE.

\subsection{Summary of Results}
\subsubsection{MAC IE}
\begin{figure}
\begin{center}
\psfrag{Encoder 1}{Encoder~1}
\psfrag{Encoder 2}{Encoder~2}
\psfrag{Host source 1}{Host source~1}
\psfrag{Host source 2}{Host source~2}
\psfrag{Decoder}{Decoder}
\psfrag{MAC}{MAC}
\psfrag{Channel}{$p(\svy|\svx_1,\svs_1,\svx_2, \svs_2)$}
\psfrag{S1}{$\rvs_{1}^n$}
\psfrag{S2}{$\rvs_{2}^n$}
\psfrag{X1}{$\rvx_{1}^n$}
\psfrag{X2}{$\rvx_{2}^n$}
\psfrag{Y}{$\rvy^n$}
\psfrag{W1}{$\rvw_{1}$}
\psfrag{W2}{$\rvw_{2}$}
\psfrag{DECOUT}{$g(\rvy^n)$}
\resizebox{.8\columnwidth}{!}{
\includegraphics{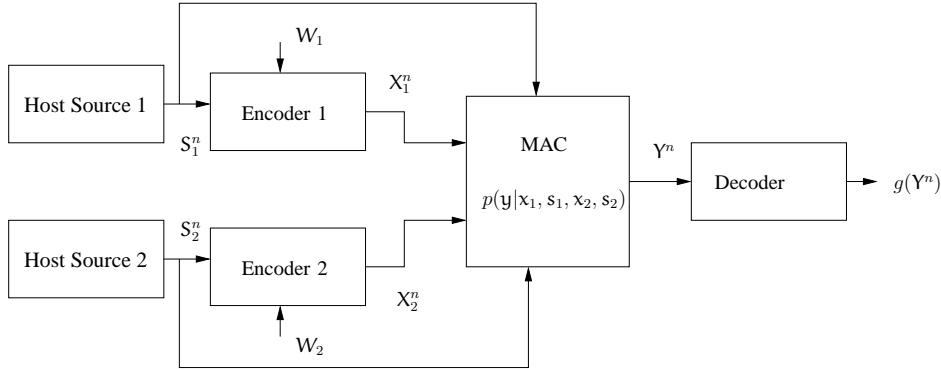}
}
\end{center}
\caption{ \label{fig:distributedRIE}  Block diagram of multiple access channel 
information embedding model.}
\end{figure}

In Section~\ref{sec:distributedRIE}, we consider  a 
two-user MAC IE model shown in Figure~\ref{fig:distributedRIE},
but the results can be extended to any number of users.
Encoder $i$ embeds its information $\rvw_i$ into  a host signal
$\rvs_i^n$, generated by a host source $i$, such that
the per-letter distortion between $\rvs_i^n$ and
$\rvx_i^n$ is less than $\Delta_i$, $i=1,2$.

For this model, we consider the following three cases in recovering,
in the sense of probability of error going to zero, the messages and the
host sequences at the decoder from the received sequence $\rvy^n$:
\begin{itemize}
\item \textbf{Case~A, Recovery of Neither Host:} The decoder recovers
$(\rvw_1,\rvw_2)$  from $\rvy^n$.
\item \textbf{Case~B, Recovery of One Host:} The decoder recovers
$(\rvw_1,\rvw_2)$ along with the one host from $\rvy^n$.
Without loss of generality,
we can assume that the host sequence $\rvs_2^n$ of Encoder~2 
is recovered at the decoder.
\item \textbf{Case~C, Recovery of Both Hosts :} The decoder recovers
$(\rvw_1,\rvw_2)$ and  $(\rvs_1^n, \rvs_2^n)$ from $\rvy^n$.
\end{itemize}
Our general MAC IE model considers scenarios in which
the MAC output potentially depends on both the embedded signals and  the host
signals. For Cases A and B,  we develop inner bounds on the 
respective capacity regions in Sections \ref{sec:mac_nohost} 
and \ref{sec:mac_onehost}, respectively.
For Case C, we derive inner and outer bounds
on the capacity region if the hosts are correlated 
in Section~\ref{sec:mac_bothhost}, and 
we show that there is no gap between the inner and the outer bounds 
if  the hosts are independent. 


\subsubsection{BC IE}
\begin{figure}[h]
\begin{center}
\psfrag{E}[cc]{Encoder}
\psfrag{H}[cc]{Host source}
\psfrag{D2}[cc]{Decoder 2}
\psfrag{D1}[cc]{Decoder 1}
\psfrag{BC}[cc]{Broadcast Channel}
\psfrag{P1}[cc]{$p(\svy,\svz|\svx,\svs)$}
\psfrag{S}{$\rvs^n$}
\psfrag{X}{$\rvx^n$}
\psfrag{Y1}{$\rvy^n$}
\psfrag{Y2}{$\rvz^n$}
\psfrag{D}{$\mathbb{E}d(\rvs^n,\rvx^n)\leq \Delta$}
\psfrag{(w1,w2)}{$(\rvw_{1},\rvw_2)$}
\psfrag{DEC1_OUT}{$g_1(\rvy^n)$}
\psfrag{DEC2_OUT}{$g_2(\rvz^n)$}
\resizebox{.9 \columnwidth}{!}{
\includegraphics{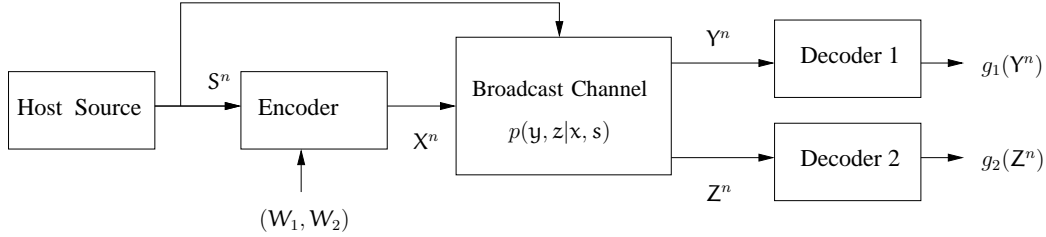} }
\end{center}
\caption{ \label{fig:broadcastRIE}  
Block diagram of the broadcast information embedding model.}
\end{figure}

In Section~\ref{sec:broadcastRIE}, we consider IE in a 
broadcast scenario as shown in
Figure~\ref{fig:broadcastRIE}, which illustrates only two decoders; in
principle the model and results can be extended to any number of
decoders. In this model, the encoder embeds two independent messages
$(\rvw_1,\rvw_2)$ into a single host sequence $\rvs^n$ 
such that the distortion between the embedded signal $\rvx^n$
and $\rvs^n$ satisfies a given distortion constraint
$\Delta$. In this paper, we focus on the case of a degraded broadcast channel, i.e., 
$p(\svy,\svz|\svx,\svs)=p(\svy|\svx,\svs)p(\svz|\svy).$ 
Decoder~1, or the \textit{better decoder}, receives the channel
output $\rvy^n$ which is drawn according to a memoryless probability law 
$p(\svy|\svx,\svs)$ for given $\rvx^n$ and $\rvs^n$.
Decoder~2, or the \textit{worse decoder}, receives the
sequence $\rvz^n$ which is  corrupted version of $\rvy^n$. 

For this model, we consider the following four cases in recovering,
in the sense of probability of error going to zero, the messages and the
host sequences at the decoders:
\begin{itemize}
\item \textbf{Case~$A'$, No Host Recovery: } Decoder~1 recovers
$(\rvw_1,\rvw_2)$  from $\rvy^n$;   Decoder~2 recovers $\rvw_2$ from $\rvz^n$.
\item \textbf{Case~$B'$, Host Recovery at the Better Decoder:} Decoder~1 recovers
$(\rvw_1,\rvw_2)$ and $\rvs^n$  from $\rvy^n$; Decoder~2 recovers $\rvw_2$ from $\rvz^n$.
\item \textbf{Case~$C'$, Host Recovery at Both Decoders:} Decoder~1 recovers
$(\rvw_1,\rvw_2)$ and  $\rvs^n$ from $\rvy^n$; Decoder~2 recovers $\rvw_2$ and
$\rvs^n$ from $\rvz^n$.
\item \textbf{Case~$D'$, Host Recovery at the Worse Decoder:} Decoder~1 recovers
$(\rvw_1,\rvw_2)$ from $\rvy^n$;  Decoder~2 recovers $\rvw_2$  and $\rvs^n$ from $\rvz^n$.
\end{itemize}
Inner and outer bounds for the BC IE capacity region in Case~$A'$
{\it{without}} an encoder distortion constraint are derived in
\cite{ysteinberg05:it}; in this paper, we extend the results to
incorporate an encoder distortion constraint in Section~\ref{sec:caseA}. 
For Case~$B'$, we develop
inner and outer bounds for the BC IE 
capacity region in Section~\ref{sec:caseB}, and for
cases~$C'$ and $D'$ we derive the BC IE 
capacity region in Section~\ref{sec:caseC} and
Section~\ref{sec:caseD}, respectively. It turns out
that the capacity regions in Cases~$C'$ and 
$D'$ are identical because
the channel output $\rvz^n$ is a degraded 
version of $\rvy^n$.
The capacity region  for the model considered 
in Case~$C'$ if compressed hosts
are available at the decoders is obtained 
in \cite{steinberg06:pisit}.

\subsection{Notation}
Throughout the paper, random variables and sample values are denoted
in a special font, e.g.,  random variable $\rvx$ and sample value
$\svx$.  Alphabets are denoted in calligraphic font, e.g., $\calX$, and
are all discrete. The shorthand $\rvx_1^n$ represents the sequence
$\rvx_{1,1},\rvx_{1,2},\ldots,\rvx_{1,n}$, and $\rvx_{1,i}^n$
represents the sequence $\rvx_{1,i},\rvx_{1,i+1},\ldots,\rvx_{1,n}$.
Finally, $\mathbb{H}(\cdot)$ and $\mathbb{I}(\cdot;\cdot)$ denote the
standard information-theoretic quantities of (ensemble average)
entropy and mutual information, respectively.


\section{MAC IE } \label{sec:distributedRIE}
In this section, let us formally discuss the model shown in
Figure~\ref{fig:distributedRIE}.
Host source $i$ generates a sequence $\rvs_i^n=\rvs_{i1}\rvs_{i2}\ldots \rvs_{in}$
of symbols from the discrete alphabet $\calS_i$, $i=1,2$.
We assume that the host sequence pair $(\rvs_1^n,\rvs_2^n)$ is generated by repeated
independent drawings of a pair of discrete random variables
$(\rvs_1,\rvs_2)$
from a given joint distribution $p(\svs_1,\svs_2)$. The host sequence
$\rvs_i^n$ is non-causally known at Encoder
$i$ for $i=1,2.$ The message source at
Encoder~$i$ produces the message index $\rvw_i \in \calW_i=\{1,2,\ldots,M_i\}$
with equal probability $1/M_i$, for $i=1,2$.
The message index at any encoder is independent of all host sequences
and also independent of the messages at all other encoders.
The rate at Encoder $i$, in bits per channel use,
is defined as $R_i=(1/n)\log_2(M_i).$

\begin{definition}
A $(M_1,M_2, D_1^{(n)}, D_2^{(n)}, n)$  \textit{MAC IE
code} consists of sequences of encoding functions at
Encoder 1 and Encoder 2,
\[ f_1^n : \mathcal{W}_1 \times \calS_1^n \rightarrow \mathcal{X}_1^n,~~\mathrm{and}~~
f_2^n : \mathcal{W}_2 \times
\calS_2^n \rightarrow \mathcal{X}_2^n,\]
respectively, and a sequence of decoding functions,
\begin{itemize}
\item \textbf{Recovery of Neither Host} $g_A^n: \mathcal{Y}^n \rightarrow (\calW_1,\mathcal{W}_2)$
\item \textbf{Recovery of One Host} $g_B^n: \mathcal{Y}^n
\rightarrow (\calW_1,\mathcal{W}_2,\mathcal{S}_2^n)$
\item \textbf{Recovery of Both Hosts} $g_C^n: \mathcal{Y}^n
\rightarrow (\calW_1,\mathcal{S}_1^n,\mathcal{W}_2,\mathcal{S}_2^n)$
\end{itemize}
The distortions associated with MAC IE code are defined as
$ D_i^{(n)} = \mathbb{E}d_i(\rvs_i^n,\rvx_i^n)$
for the additive distortion function 
\[d_i(\rvs_i^n,\rvx_i^n)= \frac{1}{n}\sum_{j=1}^{n}d_i(\rvs_{ij},\rvx_{ij})\]
for some non-negative bounded distortion functions $d_i(\rvs_{ij},\rvx_{ij})$,
where $i=1,2.$
\end{definition}

The embedded signals $\rvx_1^n$ and $\rvx_2^n$ from Encoder 1 and Encoder 2,
respectively are transmitted across a MAC
$p(\svy|\svx_1, \svs_1,\svx_2,\svs_2)$ without feedback modeled as a
memoryless conditional probability distribution
\begin{equation}
\mathrm{Pr}(\svy^n|\svx_1^n,\svs_1^n,\svx_2^n,\svs_2^n)=
\prod_{j=1}^{n}p(\svy_j|\svx_{1j},\svs_{1j},\svx_{2j},\svs_{2j}).
\label{eqn:macchannellaw}
\end{equation}

\begin{definition}
A rate pair $(R_1,R_2)$ for a given
distortion pair $(\Delta_1,\Delta_2)$ is said to be \textit{MAC IE achievable}
if there exists a sequence of
$(\lceil2^{nR_1}\rceil, \lceil2^{nR_2}\rceil, D_1^{(n)}, D_2^{(n)}, n)$
MAC IE codes with $\lim_{n \rightarrow \infty}
D_i^{(n)} \leq \Delta_i$, for $i=1,2$, and
$\lim_{n \rightarrow \infty} P_e^n=0,$ where $P_e^n$ is the probability of error defined
appropriately for each case in the sequel of this section.
\end{definition}


\begin{definition} \label{def:ib_pdfconstraints}
For given $p(\svs_1,\svs_2)$ and $p(\svy|\svx_1,\svs_1,\svx_2,\svs_2)$,
let $\mathcal{P}^i_{\mathrm{\mathrm{MAC}}}(\Delta_1,\Delta_2)$ be the  set of
all random variable tuples $(\rvq,\rvs_1,\rvs_2, (\rvu_1,\rvx_1),
(\rvu_2,\rvx_2),\rvy)$ taking values 
in finite alphabets $\mathcal{Q}$, $\mathcal{S}$, $\calU_1 \times \mathcal{X}_1$,
$\calU_2 \times \mathcal{X}_2$, and $\calY$, respectively, with joint distribution 
satisfying conditions
\begin{enumerate}
\item[a)] $\sum_{\svq,(\svu_1,\svx_1),(\svu_2,\svx_2),\svy}p(\svq,\svs_1,\svs_2,
(\svu_1,\svx_1),(\svu_2,\svx_2),\svy)=p(\svs_1,\svs_2),$
\item[b)] $p(\svq,\svs_1,\svs_2,(\svu_1,\svx_1),(\svu_2,\svx_2),\svy)=
p(\svq)p(\svs_1,\svs_2)p(\svu_1,\svx_1|\svs_1,\svq)p(\svu_2,\svx_2|\svs_2,\svq)
p(\svy|\svx_1,\svs_1,\svx_2,\svs_2)$
\item[c)] $\mathbb{E}d_i(\rvs_i,\rvx_i)\leq\Delta_i$, for $i=1,2$.
\end{enumerate}
\end{definition}

\begin{definition} \label{def:ob_pdfconstraints}
For given $p(\svs_1,\svs_2)$ and $p(\svy|\svx_1,\svs_1,\svx_2,\svx_2)$,
let $\mathcal{P}^o_{\mathrm{MAC}}(\Delta_1,\Delta_2)$ be the  set of
all random variable tuples $(\rvq,\rvs_1,\rvs_2,\rvx_1,\rvx_2,\rvy)$
taking values in finite alphabets $\mathcal{Q}$, $\mathcal{S}$, $\mathcal{X}_1$,
$\mathcal{X}_2$, and $\calY$, respectively, with joint
distribution satisfying the conditions
\begin{enumerate}
\item[a).] $\sum_{\svq,\svx_1,\svx_2,\svy}
p(\svq,\svs_1,\svs_2,\svx_1,\svx_2,\svy)=p(\svs_1,\svs_2),$
\item[b).]
$p(\svq,\svs_1,\svs_2,\svx_1,\svx_2,\svy)=p(\svq)p(\svs_1,\svs_2)
p(\svx_1,\svx_2|\svs_1,\svs_2,\svq)p(\svy|\svx_1,\svs_1,\svx_2,\svs_2),$
\item[c).] $\mathbb{E}d_i(\rvs_i,\rvx_i)\leq\Delta_i$, for $i=1,2$.
\end{enumerate}
\end{definition}

\subsection{Recovery of Neither Host}
\label{sec:mac_nohost}
In this section, we derive an inner bound on the MAC IE
capacity region for Case~A, in which the decoder recovers only
$(\rvw_1,\rvw_2)$  from $\rvy^n$.  We define the MAC  IE capacity
region $\calC_\mathrm{MAC,A}(\Delta_1,\Delta_2)$ 
as the closure of the set of all MAC IE achievable rates
$(R_1,R_2)$ with $P_{e}^{(n)}:=\mathbb{P}[(g_A^n(\rvy^n)\neq
(\rvw_1,\rvw_2)] \rightarrow 0$ as $n \rightarrow \infty$. 
The following theorem provides an inner bound on
the capacity region.

\begin{proposition}\label{prop:inner_mac_nohost}
Let $\calR_{\mathrm{MAC,A}}^{\mathrm{i}}(\Delta_1,\Delta_2)$ be the 
closure of the set of all rate pairs
$(R_1,R_2)$ such that
\begin{subequations} \label{eqn:innerbound_mac_nohost}
\begin{align}
R_1 &\leq \mathbb{I}(\rvu_1;\rvu_2,\rvy|\rvq)-\mathbb{I}(\rvu_1;\rvs_1|\rvq),\\
R_2 &\leq \mathbb{I}(\rvu_2;\rvu_1,\rvy|\rvq)-\mathbb{I}(\rvu_2;\rvs_2|\rvq), \\
R_1+R_2 & \leq \mathbb{I}(\rvu_1,\rvu_2;\rvy|\rvq)-\mathbb{I}(\rvu_1, \rvu_2;\rvs_1,\rvs_2|\rvq)
\end{align}
\end{subequations}
for some $(\rvq,\rvs_1,\rvs_2, (\rvu_1,\rvx_1),(\rvu_2,\rvx_2),\rvy) \in 
\calP_{\mathrm{\mathrm{MAC}}}^i(\Delta_1,\Delta_2)$, where
$\rvu_1$ and $\rvu_2$ are auxiliary random variables.
Then,  $\calR_{\mathrm{MAC,A}}^{\mathrm{i}}(\Delta) \subseteq \calC_{\mathrm{MAC,A}}.$
\end{proposition}

\noindent \textit{Remarks }
\begin{itemize}
\item The inner bound in
Proposition~\ref{prop:inner_mac_nohost} is similar
to that in \cite{sun04:ih}, which considers a Gaussian MAC with no host recovery, but
the result here is for the discrete memoryless case.
Because the coding procedures, and error events
in \cite{sun04:ih} apply, we do not provide a proof here.
\item To achieve the  inner bound,
distortion-constrained Gel'fand-Pinsker codes can be used to embed $\rvw_1$ and $\rvw_2$
into the host sequences $\rvs_1^n$ and $\rvs_2^n$ such that the distortion constraints 
$\Delta_1$ and $\Delta_2$ are met, respectively.
\end{itemize}


\subsection{Recovery of One Host}
\label{sec:mac_onehost}
In this section, we derive inner and outer bounds on the MAC IE
capacity region for Case~B, in which the decoder recovers
$(\rvw_1,\rvw_2,\rvs_2^n)$  from $\rvy^n$.  We define the MAC  IE capacity
region $\calC_\mathrm{MAC,B}(\Delta_1,\Delta_2)$ as the closure of the set of all 
MAC IE achievable rates $(R_1,R_2)$ with $P_{e}^{(n)}:=\mathbb{P}[(g_B^n(\rvy^n)\neq
(\rvw_1,\rvw_2,\rvs_2^n)] \rightarrow 0$ as $n \rightarrow \infty$.
The following theorem provides an inner bound for
the capacity region.

\begin{proposition}\label{prop:inner_mac_onehost}
Let $\calR_{\mathrm{MAC,B}}^{\mathrm{i}}(\Delta_1,\Delta_2)$
be the closure of the set of all rate pairs
$(R_1,R_2)$ such that
\begin{subequations} \label{eqn:inner_mac_onehost}
\begin{align}
R_1 &\leq \mathbb{I}(\rvu_1;\rvy|\rvx_2,\rvs_2,\rvq)-\mathbb{I}(\rvu_1;\rvs_1|\rvx_2,\rvs_2,\rvq),\\
R_2 &\leq \mathbb{I}(\rvx_2,\rvs_2;\rvy|\rvu_1,\rvq)-\mathbb{H}(\rvs_2|\rvu_1,\rvq), \\
R_1+R_2 & \leq
\mathbb{I}(\rvu_1,\rvx_2,\rvs_2;\rvy|\rvq)-\mathbb{H}(\rvs_2)-\mathbb{I}(\rvu_1;\rvs_1|\rvx_2,\rvs_2,\rvq)
\end{align}
\end{subequations}
for some $(\rvq,\rvs_1,\rvs_2, (\rvu_1,\rvx_1),(\rvx_2,\rvx_2),\rvy) \in \calP_{\mathrm{MAC}}^i(\Delta_1,\Delta_2)$, where
$\rvu_1$ and $\rvq$ are auxiliary random variables.
Then,  $\calR_{\mathrm{MAC,B}}^{\mathrm{i}}(\Delta_1,\Delta_2) \subseteq \calC_{\mathrm{MAC,B}}(\Delta_1,\Delta_2)$
\end{proposition}

\noindent \textit{Remarks}
\begin{itemize}
\item The inner bound in
Proposition~\ref{prop:inner_mac_onehost} is a special case of an inner bound in
 \cite{kotagiri06:it}, which considers the
state-dependent MAC with state known at one encoder 
and recovery of only messages at the decoder.
To obtain the inner bound in Proposition~\ref{prop:inner_mac_onehost},
substitute $(\rvx_2,\rvs_2)$ in place of $\rvx_2$ into the inner bound in \cite{kotagiri06:it}.
\item To achieve the  inner bound,
distortion constrained  Gel'fand-Pinsker coding is used to embed $\rvw_1$
into the host sequence $\rvs_1^n$, and
distortion-constrained superposition coding is used to embed $\rvw_2$
into the host sequence $\rvs_2^n$.
\item If we choose $\rvu_2=(\rvx_2,\rvs_2)$ int Proposition~\ref{prop:inner_mac_nohost},
we obtain the inner bound in Proposition~\ref{prop:inner_mac_onehost}. Thus, 
$\calR_{\mathrm{MAC,B}}^{\mathrm{i}}(\Delta_1,\Delta_2) \subseteq \calR_{\mathrm{MAC,A}}^{\mathrm{i}}(\Delta_1,\Delta_2).$
\end{itemize}

%



\subsection{Recovery of Both Hosts}
\label{sec:mac_bothhost}

In this section, we derive inner and outer bounds on the MAC IE
capacity region for Case~C in which the decoder recovers
$(\rvw_1,\rvs_1^n,\rvw_2,\rvs_2^n)$  from $\rvy^n$.  We define the MAC  IE capacity
region $\calC_\mathrm{MAC,C}(\Delta_1,\Delta_2)$ as the closure of all MAC IE achievable rates
$(R_1,R_2)$ with $P_{e}^{(n)}:=\mathbb{P}[(g(\rvy^n)\neq
(\rvw_1,\rvs_1^n,\rvw_2,\rvs_2^n)] \rightarrow 0$ as $n \rightarrow \infty$.
The following theorem obtains an inner bound for
the capacity region.

\begin{theorem} \label{thm:innerbound_distributedRIE}
Let $\calR_{\mathrm{MAC,C}}^{\mathrm{i}}(\Delta_1,\Delta_2)$ be the set of all rate pairs
$(R_1,R_2)$ such that
\begin{subequations} \label{eqn:innerbound_distributedRIE}
\begin{align}
R_1  & <  [\mathbb{I}(\rvx_1,\rvs_1;\rvy|\rvx_2,\rvs_2,\rvq)-
\mathbb{H}(\rvs_1|\rvs_2)], \label{eqn:innerbound_distributedRIE_R1} \\
R_2  & < [\mathbb{I}(\rvx_2,\rvs_2;\rvy|\rvx_1,\rvs_1,\rvq)-
\mathbb{H}(\rvs_2|\rvs_1)], \label{eqn:innerbound_distributedRIE_R2} \\
R_1+R_2 & < [\mathbb{I}(\rvx_1,\rvs_1,\rvx_2,\rvs_2;\rvy|\rvq)-
\mathbb{H}(\rvs_1,\rvs_2)], \label{eqn:innerbound_distributedRIE_R3}
\end{align}
\end{subequations}
for some $(\rvq,\rvs_1,\rvs_2,(\rvx_1,\rvx_1),(\rvx_2,\rvx_2),\rvy)\in 
\mathcal{P}^i_{\mathrm{\mathrm{MAC}}}(\Delta_1,\Delta_2)$.
Then, $$\calR_{\mathrm{MAC,C}}^{\mathrm{i}}(\Delta_1,\Delta_2)
\subseteq \calC_{\mathrm{MAC,C}}(\Delta_1,\Delta_2).$$
\end{theorem}

\textbf{Proof:} See Appendix~\ref{sec:proof_innerbound_DRIE}

The following theorem gives an outer bound for the capacity region 
if $\rvs_1$ and $\rvs_2$ are correlated.
\begin{theorem} \label{thm:outerbound_distributedRIE}
Let $\calR_{\mathrm{MAC,C}}^{\mathrm{o}}(\Delta_1,\Delta_2)$ 
be the set of all rate pairs $(R_1,R_2)$ such that
\begin{subequations} \label{eqn:outerbound_DRIE}
\begin{align}
R_1 & < [\mathbb{I}(\rvx_1,\rvs_1;\rvy|\rvx_2,\rvs_2,\rvq)-
\mathbb{H}(\rvs_1|\rvs_2)],
\label{eqn:outerbound_DRIE_R1} \\
R_2 & < [\mathbb{I}(\rvx_2,\rvs_2;\rvy|\rvx_1,\rvs_1,\rvq)-
\mathbb{H}(\rvs_2|\rvs_1)],
\label{eqn:outerbound_DRIE_R2} \\
R_1+R_2 &  < [\mathbb{I}(\rvx_1,\rvs_1,\rvx_2,\rvs_2;\rvy|\rvq)-
\mathbb{H}(\rvs_1,\rvs_2)],
\label{eqn:outerbound_DRIE_R3}
\end{align}
\end{subequations}
for some $(\rvq,\rvs_1,\rvs_2,\rvx_1,\rvx_2,\rvy) \in 
\mathcal{P}^o_{\mathrm{MAC}}(\Delta_1,\Delta_2)$.
If the host random variables $\rvs_1$ and $\rvs_2$ are correlated,
then \[\calC_{\mathrm{MAC,C}}(\Delta_1,\Delta_2)
\subseteq \calR_{\mathrm{MAC,C}}^{\mathrm{o}}(\Delta_1,\Delta_2).\]
If the host random variables $\rvs_1$ and $\rvs_2$ are independent,
then \[\calC_{\mathrm{MAC,C}}(\Delta_1,\Delta_2)
\subseteq \calR_{\mathrm{MAC,C}}^{\mathrm{i}}(\Delta_1,\Delta_2).\]
\end{theorem}

\textbf{Proof:} See Appendix~\ref{sec:proof_outerbound_DRIE}

The following corollary of Theorem~\ref{thm:innerbound_distributedRIE} and 
Theorem~\ref{thm:outerbound_distributedRIE} states the MAC IE
capacity region for a given pair of distortion constraints
$(\Delta_1,\Delta_2)$ if the host random variables
$\rvs_1$ and $\rvs_2$ are independent.

\begin{corollary} \label{thm:distributedRIE_capacity}
If the host random variables $\rvs_1$ and $\rvs_2$ are independent, then
the capacity region $\calC_{\mathrm{MAC,C}}(\Delta_1,\Delta_2)$ is the closure of
the set of all rate pairs
$(R_1,R_2)$ such that
\begin{subequations}\label{eqn:distributedRIE_capacity}
\begin{align}
R_1 &  < [\mathbb{I}(\rvx_1,\rvs_1;\rvy|\rvx_2,\rvs_2,\rvq)-
\mathbb{H}(\rvs_1|\rvs_2)], \label{eqn:distributedRIE_capacity_R1} \\
R_2 & < [\mathbb{I}(\rvx_2,\rvs_2;\rvy|\rvx_1,\rvs_1,\rvq)-
\mathbb{H}(\rvs_2|\rvs_1)], \label{eqn:distributedRIE_capacity_R2} \\
R_1+R_2 &  < [\mathbb{I}(\rvx_1,\rvs_1,\rvx_2,\rvs_2;\rvy|\rvq)-
\mathbb{H}(\rvs_1,\rvs_2)], \label{eqn:distributedRIE_capacity_R3}
\end{align}
\end{subequations}
for some $(\rvq,\rvs_1,\rvs_2,(\rvx_1,\rvx_1),(\rvx_2,\rvx_2),\rvy)\in
\mathcal{P}^i_\mathrm{MAC}(\Delta_1,\Delta_2)$.
\end{corollary}

\noindent \textit{Remarks}
\begin{itemize}
\item  To compute either (\ref{eqn:innerbound_distributedRIE}) or (\ref{eqn:outerbound_DRIE}), 
it is sufficient to consider time-sharing random variable $\rvq$ with $|\calQ| \leq 4$
by Caratheodory's theorem \cite{cover91:book}.
\item In most communication scenarios, message transmission rates of zero 
are achievable. However, in this model, message transmission rates of zero 
can be  unachievable if the host source pair
$p(\svs_1,\svs_2)$ is such that the upper bounds on
$R_1$, $R_2$ and $R_1+R_2$ in (\ref{eqn:distributedRIE_capacity}) are negative.
This is because we require host recovery at the decoder as well.
\end{itemize}

\section{Degraded BC IE}
\label{sec:broadcastRIE}
In this section, let us formally define the BC IE model shown
in Figure~\ref{fig:broadcastRIE}.
A host sequence $\rvs^n=(\rvs_1,\rvs_2,\ldots,\rvs_n)$ is an
independent and identically distributed (i.i.d.) discrete random
sequence whose elements are drawn with probability mass function
$p(\svs)$, $\svs\in \calS$. All alphabets are
discrete. We assume that the host sequence $\rvs^n$ is non-causally
known at the encoder.  The encoder embeds a message pair
$(\rvw_1,\rvw_2)$ into the host sequence $\rvs^n$ such that the
average distortion between $\rvs^n$ and the embedded sequence
$\rvx^n$ satisfies a given distortion constraint $\Delta$.  The messages
$\rvw_1 \in \{1,2,\ldots,M_1\}$ and $\rvw_2 \in \{1,2,\ldots,M_2\}$
are drawn equally likely with probabilities $1/M_1$ and $1/M_2$,
respectively.  Then the rate of message $\rvw_i$ is given by
$R_i=(1/n)\log_2M_i$ bits per channel use, for $i=1,2$. It is also
assumed that the message $\rvw_i$ is independent of the other message
and the host sequence for $i=1,2.$

\begin{definition}
A $(M_1, M_2, D^{(n)},n)$ \textit{BC IE
code} consists of a sequence of encoding functions at the encoder
$$f^n: \mathcal{W}_1 \times \calW_2 \times \calS^n \rightarrow
\mathcal{X}^n,$$
and a sequence of decoding functions at Decoder~1 and Decoder~2
\begin{itemize}
\item \textbf{No Host Recovery}~~$g_{1,A'}^n: \mathcal{Y}^n \rightarrow (\calW_1,\calW_2)
~~\mathrm{and}~~ g_{2,A'}^n: \mathcal{Z}^n \rightarrow \calW_2$
\item \textbf{Host Recovery at the Better Decoder}~~ $g_{1,B'}^n: \mathcal{Y}^n \rightarrow (\calW_1,\calW_2,\calS^n)
~~\mathrm{and}~~ g_{2,B'}^n: \mathcal{Z}^n \rightarrow \calW_2$
\item \textbf{Host Recovery at Both Decoders}~~$g_{1,C'}^n: \mathcal{Y}^n \rightarrow (\calW_1,\calW_2,\calS^n)
~~\mathrm{and}~~ g_{2,C'}^n: \mathcal{Z}^n \rightarrow (\calW_2,\calS^n)$
\item \textbf{Host Recovery at the Worse Decoder}~~$g_{1,D'}^n: \mathcal{Y}^n \rightarrow (\calW_1,\calW_2)
~~\mathrm{and}~~ g_{2,D'}^n: \mathcal{Z}^n \rightarrow (\calW_2,\calS^n),$
\end{itemize}
respectively. The associated distortion is defined as $
D^{(n)} = \mathbb{E}d(\rvs^n,\rvx^n),$ where $d(\rvs^n,\rvx^n)=
(1/n)\sum_{j=1}^{n}d(\rvs_{j},\rvx_{j})$ for given non-negative bounded
distortion measure $d(\cdot,\cdot)$.
\end{definition}

The embedded signal $\rvx^n$ is transmitted across a discrete
memoryless degraded broadcast channel (DMDBC) with state,
$p(y|x,s)p(z|y)$, modeled as a memoryless conditional probability
distribution
\begin{equation}
\mathrm{Pr}(\rvy^n=\svy^n,\rvz^n=\svz^n|\svx^n,\svs^n)=
\prod_{j=1}^{n}p(\svy_{j}|\svx_{j},\svs_{j})p(\svz_{j}|\svy_{j}).
\label{eqn:broadcastchannellaw}
\end{equation}

\begin{definition}
A rate pair $(R_1,R_2)$ for a given distortion $\Delta$
is said to be \textit{BC IE achievable} if there exists a sequence of
$(\lceil2^{nR_1}\rceil,\lceil2^{nR_2}\rceil,D^{(n)},n)$ BC IE
codes with $\lim_{n \rightarrow \infty} D^{(n)}
\leq \Delta$ and $\lim_{n \rightarrow \infty} P_e^n=0,$ where $P_e^n$
is the probability of error defined appropriately for each case in the
sequel of the paper.
\end{definition}

\begin{definition}
For a given $p(\svs)$ and $p(\svy|\svx,\svs)p(\svz|\svy)$, let $\calP(\Delta)$ be the
collection of random variables $(\rvt,\rvs,\rvx,\rvy,\rvz)$ with joint
probability mass function satisfying the following conditions
\begin{itemize}
\item[a)] $p(\svt,\svs,\svx,\svy,\svz) = p(\svt,\svs,\svx)p(\svy|\svx,\svs)p(\svz|\svy)$
\item[b)] $\sum_{\svt \in \calT, \svx \in \calX} p(\svt,\svx,\svs)=p(\svs)$
\item[c)] $\mathbb{E}d(\rvs,\rvx) \leq \Delta, $
\end{itemize}
where $\rvt$ is an auxiliary random variable.
\end{definition}

\subsection{No Host Recovery}\label{sec:caseA}
In this section, we state inner and outer bounds for the BC IE
capacity region in Case~$A'$, in which Decoder~1 recovers
$(\rvw_1,\rvw_2)$ from $\rvy^n$ and Decoder~2 recovers $\rvw_2$ from
$\rvz^n$. The BC IE capacity region $\calC_{A'}(\Delta)$ is the
closure of all BC IE achievable rates $(R_1,R_2)$ with
$P_{e}^{(n)}:=\mathrm{Pr}[(g_{1,A'}^n(\rvy^n)\neq(\rvw_1,\rvw_2)~
\mathrm{or}~g_{2,A'}^n(\rvz^n)\neq
\rvw_2] \rightarrow 0$ as $n \rightarrow \infty$.

\begin{proposition}\label{prop:innerandouter_schemeA}
Let $\calR_{A'}^i(\Delta)$ be the closure of the set of all rate pairs
$(R_1,R_2)$ such that
\begin{subequations} \label{innerbound_schemeA}
\begin{align}
R_1 &\leq \mathbb{I}(\rvv;\rvy|\rvu)-\mathbb{I}(\rvv;\rvs|\rvu),\\
R_2 &\leq \mathbb{I}(\rvu;\rvz) -\mathbb{I}(\rvu;\rvs),
\end{align}
\end{subequations}
for some $((\rvu,\rvv),\rvs,\rvx,\rvy,\rvz) \in \calP(\Delta)$, where
$\rvu$ and $\rvv$ are auxiliary random variables with alphabet sizes
satisfying $|\calU| \leq |\calX||\calS|+1$ and $|\calV| \leq
|\calX||\calS|(|\calX||\calS|+1)$, respectively.  Let
$\calR_{A'}^o(\Delta)$ be the closure of the set of all rate pairs
$(R_1,R_2)$ such that
\begin{subequations} \label{outerbound_schemeA}
\begin{align}
R_1 &\leq \mathbb{I}(\rvv;\rvy|\rvu,\rvw)-\mathbb{I}(\rvv;\rvs|\rvu,\rvw), \\
R_2 &\leq \mathbb{I}(\rvu;\rvz) -\mathbb{I}(\rvu;\rvs),  \\
R_1+R_2 &\leq \mathbb{I}(\rvu,\rvv,\rvw;\rvy) - \mathbb{I}(\rvu,\rvv,\rvw;\rvs),
\end{align}
\end{subequations}
for some $((\rvu,\rvv,\rvw),\rvs,\rvx,\rvy,\rvz) \in \calP(\Delta)$,
where $\rvu$, $\rvw$, and $\rvw$ are auxiliary random variables with
alphabet sizes satisfying $|\calU| \leq |\calX||\calS|+2$, $|\calV|
\leq |\calX||\calS|(|\calX||\calS|+2)+1$, and $\calW \leq
(|\calX||\calS|(|\calX||\calS|+2)+1)(|\calX||\calS|+2)|\calX||\calS|+1$,
respectively.  Then, $\calR^i_{A'}(\Delta) \subseteq \calC_{A'}(\Delta)
\subseteq \calR^o_{A'}(\Delta)$.
\end{proposition}

\noindent \textit{Remarks} \\
The inner and outer bounds in
Proposition~\ref{prop:innerandouter_schemeA} are slightly different
from those in \cite{ysteinberg05:it}, which does not consider an
encoder distortion constraint.  Although essentially the same proofs
in \cite{ysteinberg05:it} apply, here there is an additional
constraint on the joint probability mass functions $\calP(\Delta)$ to
limit the average distortion between the host $\rvs$ and the channel
input $\rvx$ to be at most $\Delta$.  To achieve the  inner bound,
Gel'fand-Pinsker codes can be used to embed the messages
$(\rvw_1,\rvw_2)$ into the host sequence $\rvs^n$.

\subsection{Host Recovery at the Better Decoder} \label{sec:caseB}
In this section, we derive inner and outer bounds on the BC IE
capacity region in Case~$B'$, in which Decoder~1 recovers
$(\rvw_1,\rvw_2)$ and $\rvs^n$ from $\rvy^n$ and Decoder~2 recovers
only $\rvw_2$ from $\rvz^n$. We define the BC IE capacity
region $\calC_{B'}(\Delta)$ as the closure of all BC IE achievable rates
$(R_1,R_2)$ with $P_{e}^{(n)}:=\mathrm{Pr}[(g_{1,B'}^n(\rvy^n)\neq
(\rvw_1,\rvw_2,\hat{\rvs}^n)~\mathrm{or}~ g_{2,B'}^n(\rvz^n)\neq \rvw_2]
\rightarrow 0$ as $n \rightarrow \infty$. The following two theorems give inner and
outer bounds for the capacity region in this case.
\begin{theorem} \label{thm:innerbound_schemeB}
Let $\calR_{B'}^i(\Delta)$ be the closure of the set of all rate pairs $(R_1,R_2)$ such that
\begin{subequations} \label{eqn:innerbound_schemeB}
\begin{align}
R_1 &\leq \mathbb{I}(\rvx,\rvs;\rvy|\rvu)-\mathbb{H}(\rvs|\rvu),\\
R_2 &\leq \mathbb{I}(\rvu;\rvz) -\mathbb{I}(\rvu;\rvs),
\end{align}
\end{subequations}
for some $(\rvu,\rvs,\rvx,\rvy,\rvz) \in \calP(\Delta),$ where $\rvu$
is an auxiliary random variable with alphabet size satisfying
$|\calU| \leq |\calX||\calS| + 1$.
Then $\calR_{B'}^i(\Delta) \subseteq \calC_{B'}(\Delta)$.
\end{theorem}
\noindent \textbf{Proof:} See \ref{sec:proof_innerbound_schemeB} .

\begin{theorem} \label{thm:outerbound_schemeB}
Let $\calR_{B'}^o(\Delta)$ be the closure of the set of all rate pairs
$(R_1,R_2)$ such that
\begin{subequations} \label{eqn:outerbound_schemeB}
\begin{align}
R_1 &\leq \mathbb{I}(\rvx,\rvs;\rvy|\rvu)-\mathbb{H}(\rvs|\rvu),\\
R_2 &\leq \mathbb{I}(\rvu,\rvv;\rvz) -\mathbb{I}(\rvu,\rvv;\rvs),
\end{align}
\end{subequations}
for some $((\rvu,\rvv),\rvs,\rvx,\rvy,\rvz) \in \calP(\Delta)$, where $\rvu$ and
 $\rvv$ are auxiliary random variables with alphabet sizes satisfying
 $|\calU| \leq |\calX||\calS| + 1$
and $|\calV| \leq |\calX||\calS|(|\calX||\calS| + 1)$, respectively. Then
$\calC_{B'}(\Delta) \subseteq \calR_{B'}^o(\Delta)$.
\end{theorem}
\noindent \textbf{Proof:} See Appendix~\ref{sec:proof_outerbound_schemeB}.

\noindent \textit{Remarks} \\
To obtain the above inner bound, the
message $\rvw_2$ is embedded into the host sequence $\rvs^n$ using
Gel'fand-Pinsker coding, and the message $\rvw_1$ is embedded into the
host sequence using superposition coding such that the distortion
constraint is satisfied. The above inner and outer bounds
are already convex regions. So, there is no need to introduce time-sharing
auxiliary random variables. Let us write the constraint on $R_2$ in
the outer bound given in (\ref{eqn:outerbound_schemeB}) as follows
$$ \mathbb{I}(\rvu,\rvv;\rvz) -\mathbb{I}(\rvu,\rvv;\rvs) =
\mathbb{I}(\rvu;\rvz) -\mathbb{I}(\rvu;\rvs)+ \{\mathbb{I}(\rvv;\rvz|\rvu)
-\mathbb{I}(\rvv;\rvs|\rvu)\}.$$
This term $\mathbb{I}(\rvv;\rvz|\rvu) -\mathbb{I}(\rvv;\rvs|\rvu)$ is the difference
between the inner and outer bounds. If $\rvv$ is a deterministic function of $\rvu$,
both inner and outer bounds coincide.
This clearly shows that $\calR_{B'}^i(\Delta) \subseteq \calR_{B'}^o(\Delta).$

\subsection{Host Recovery at Both Decoders}\label{sec:caseC}
This section derives the BC IE capacity region in Case~$C'$, in
which Decoder~1 recovers $(\rvw_1,\rvw_2)$ and $\rvs^n$ from $\rvy^n$
and Decoder~2 recovers $\rvw_2$ and $\rvs^n$ from $\rvz^n$. We define
the BC IE capacity region $\calC_{C'}(\Delta)$ as the closure of
all BC IE achievable rates $(R_1,R_2)$ with
$P_{e}^{(n)}:=\mathrm{Pr}[(g_{1,C'}^n(\rvy^n)\neq (\rvw_1,\rvw_2,\rvs^n)~
\mathrm{or}~ g_{2,C'}^n(\rvz^n)\neq (\rvw_2,\rvs^n)] \rightarrow 0$ as $n
\rightarrow \infty$.

\begin{theorem} \label{thm:capacity_schemeC}
$\calC_{C'}(\Delta)$ is the closure of the set of all rate pairs
$(R_1,R_2)$ such that
\begin{subequations} \label{eqn:capacity_schemeC}
\begin{align}
R_1 &\leq \mathbb{I}(\rvx;\rvy|\rvu,\rvs), \\
R_2 &\leq \mathbb{I}(\rvx,\rvs;\rvz)-\mathbb{H}(\rvs),
\end{align}
\end{subequations}
for some $(\rvu,\rvs,\rvx,\rvy,\rvz) \in \calP(\Delta)$, where $\rvu$
is an auxiliary random variable with $|\calU| \leq |\calX||\calS|$.
\end{theorem}
\noindent \textbf{Proof:} See Appendix~\ref{sec:proof_capacity_schemeC}

\noindent \textit{Remarks} \\
To achieve the BC IE capacity
region, the messages $(\rvw_1,\rvw_2)$ are embedded into the host
sequence using distortion-constrained superposition coding
as in the previous cases because lossless recovery, i.e., reversible
embedding, of the host sequence $\rvs^n$ is required in Case~$C'$.

\subsection{Host Recovery at the Worse Decoder}\label{sec:caseD}
This section derives the BC IE capacity region in Case~$D'$, in
which Decoder~1 recovers $(\rvw_1,\rvw_2)$ from $\rvy^n$ and Decoder~2
recovers $\rvw_2$ and $\rvs^n$ from $\rvz^n$. We define the broadcast
IE capacity region $\calC_{D'}(\Delta)$ as the closure of all BC IE achievable
rates $(R_1,R_2)$ with $P_{e}^{(n)}:=\mathrm{Pr}[(g_{1,D'}^n(\rvy^n)\neq
(\rvw_1,\rvw_2)~\mathrm{or}~g_{2,D'}^n(\rvz^n)\neq (\rvw_2,\rvs^n)]
\rightarrow 0$ as $n \rightarrow \infty$.
\begin{corollary}
$\calC_{D'}(\Delta) = \calC_{C'}(\Delta).$
\end{corollary}
\noindent \textbf{Proof:}
Since $\rvz^n$ is a degraded version of $\rvy^n$, and
$(\rvw_2,\rvs^n)$ must be reliably decoded from $\rvz^n$,
$(\rvw_2,\rvs^n)$ can also be decoded from $\rvy^n$. This implies that
the BC IE capacity region in Case~$D'$ is the same as in Case~$C'$.
\appendix
We present definitions related to strong
typicality~\cite{cover91:book, csiszar81:book, tung78} and
important theorems based on strong typicality which will be used
throughout the section.

 \begin{definition}
 A sequence $\svx^n \in \calX^n$ is said to be \textit{$\epsilon$-strongly
 typical} with respect to a distribution $p(\svx)$ on
 $\calX$ or $\svx^n \in T_{\epsilon}^n(\rvx)$ if
\[ \left | \frac{1}{n} N( \sva |\svx^n)-p(\sva) \right | < \frac{\epsilon}{|\calX|}, \]
for all $\sva \in \calX$ with $p(\sva)>0$, and
$N(\sva|\svx^n)=0$ for all $\sva \in \calX$ with $p(\sva)=0$, where
$N(\sva|\svx^n)$ is the number of occurrences of the symbol $\sva$ in the sequence $\rvx^n$.
 \end{definition}

\begin{definition}
A pair of sequences $(\svx^n,\svy^n) \in \calX^n\times\calY^n$
is said to be \textit{jointly $\epsilon$-strongly typical} with
respect to a distribution
$p(\svx,\svy)$ on $\calX \times \calY$  or
$(\svx^n,\svy^n) \in T_{\epsilon}^n(\svx,\svy)$ if
\[ \left | \frac{1}{n}N(\sva,\svb|\svx^n,\svy^n)-p(\sva,\svb) \right | <
\frac{\epsilon}{|\calX||\calY|}, \]
for all $(\sva,\svb) \in \calX \times \calY$ with $p(\sva,\svb)>0$,
and $N(\sva,\svb|\svx^n,\svy^n)=0$ for all
$(\sva,\svb) \in \calX \times \calY$ with
$p(\sva,\svb)=0$, where $N(\sva,\svb|\svx^n,\svy^n)$ is the number of
occurrences of the symbol $(\sva,\svb)$ in the
pair of sequences $(\svx^n,\svy^n)$.
\end{definition}

For completeness, we recall theorems on strong typicality 
\cite{cover91:book, csiszar81:book, tung78} which will be used throughout this section.
\begin{lemma} \label{thm:ST_theorem1}
Suppose $\rvx^n$ is generated from a discrete memoryless source(DMS) $p(\svx)$ and
$\rvx^n \in T_{\epsilon}^n(\rvx)$.
Then, we have the following
\begin{equation}
2^{-n[\mathbb{H}(\rvx)+\epsilon_1]}  < P^n(\svx^n) <  2^{-n[\mathbb{H}(\rvx)-\epsilon_1]}
\end{equation}
\begin{equation}
(1-\epsilon_2)\,2^{n[\mathbb{H}(\rvx)-\epsilon_1]} <  |T_{\epsilon}^n(\rvx)|  <
2^{n[\mathbb{H}(\rvx)+\epsilon_1]}
\end{equation}
\begin{equation}
(1-\epsilon_2)  \leq  \mathrm{Pr}[\rvx^n \in T_{\epsilon}^n(\rvx)] \leq  1
\end{equation}
where $\epsilon_1 \rightarrow 0$ as $\epsilon \rightarrow 0$, and
$\epsilon_2 \rightarrow 0$ as $n \rightarrow \infty$ for fixed $\epsilon$.
\end{lemma}

\begin{lemma} \label{thm:ST_lemma1}
Suppose $(\rvx^n,\rvy^n)$ is generated from a discrete memoryless source (DMS) $p(\svx,\svy)$
and $(\svx^n,\svy^n) \in T_{\epsilon}^n(\rvx,\rvy)$ and
Then, we have the following
\begin{equation}
2^{-n[\mathbb{H}(\rvx,\rvy)+\epsilon'_1]}  < P^n(\svx^n,\svy^n) <
 2^{-n[\mathbb{H}(\rvx,\rvy)-\epsilon'_1]}
\end{equation}
\begin{equation}
(1-\epsilon'_2)\,2^{n[\mathbb{H}(\rvx,\rvy)-\epsilon'_1]} <
|T_{\epsilon}^n(\rvx,\rvy)|  <  2^{n[\mathbb{H}(\rvx,\rvy)+\epsilon'_1]}
\end{equation}
\begin{equation}
(1-\epsilon'_2)  \leq  \mathrm{Pr}[(\rvx^n,\rvy^n) \in T_{\epsilon}^n(\rvx,\rvy)] \leq  1
\end{equation}
where $\epsilon'_1 \rightarrow 0$ as $\epsilon \rightarrow 0$, and
$\epsilon'_2 \rightarrow 0$ as $n \rightarrow \infty$ for fixed
$\epsilon$.
\end{lemma}

\begin{lemma} \label{thm:ST_lemma2}
Suppose $(\rvx^n,\rvy^n)$ is generated from a
discrete memoryless source(DMS) $p(\svx,\svy)$ and
$(\rvx^n,\rvy^n) \in T_{\epsilon}^n(\rvx,\rvy)$.
Then, we have the following
\begin{equation}
2^{-n[\mathbb{H}(\rvy|\rvx)+\epsilon''_1]}  < P^n(\svy^n|\svx^n) <
2^{-n[\mathbb{H}(\rvy|\rvx)-\epsilon''_1]}
\end{equation}
\begin{equation}
(1-\epsilon''_2)\,2^{n[\mathbb{H}(\rvy|\rvx)-\epsilon'_1]} <
|T_{\epsilon}^n(\rvx,\rvy|\svx^n)|  <
2^{n[\mathbb{H}(\rvy|\rvx)+\epsilon''_1]}
\end{equation}
\begin{equation}
(1-\epsilon''_2)  \leq  \mathrm{Pr}[(\svx^n,\rvy^n) \in T_{\epsilon}^n(\rvx,\rvy)] \leq  1
\end{equation}
where $\epsilon''_1 \rightarrow 0$ as $\epsilon \rightarrow 0$,
and $\epsilon''_2 \rightarrow 0$
as $n \rightarrow
\infty$ for fixed
$\epsilon$, and $T_{\epsilon}^n(\rvx,\rvy|\svx^n)=\{\svy^n:(\svx^n,\svy^n)\in
T_{\epsilon}^n(\rvx,\rvy)\}$.
\end{lemma}


\subsection{Proof of Theorem~\ref{thm:innerbound_distributedRIE} }
\label{sec:proof_innerbound_DRIE}
In this section, we demonstrate existence of a sequence of MAC IE
codes \\ $(\lceil2^{nR_1}\rceil, \lceil2^{nR_2}\rceil, D_1^{(n)}, D_2^{(n)}, n)$ with
$\lim_{n \rightarrow \infty} P_e^n=0$, and $\lim_{n \rightarrow \infty}
D_i^{(n)} \leq \Delta_i$ for $i=1,2$ if the rate pair $(R_1,R_2)$
satisfying (\ref{eqn:innerbound_distributedRIE}).
Fix  
$(\rvq,\rvs_1,\rvs_2,(\rvx_1,\rvx_1),(\rvx_2,\rvx_2),\rvy) \in 
\calP^i_{\mathrm{MAC}}(\Delta_1,\Delta_2)$ and $n$.
We construct a MAC  IE code $(\lceil2^{nR_1}\rceil, 
\lceil2^{nR_2}\rceil, D_1^{(n)}, D_2^{(n)}, n)$ as follows.
\begin{itemize}
\item \textbf{Code construction:} Throughout the achievability proof, 
let $i \in \calI =\{1,2\}.$
Generate time sharing sequence $\rvq^n =(\rvq_1,\rvq_2,
\ldots,\rvq_n)$
whose elements are i.i.d. with distribution $p(\svq)$.
At Encoder $i$, for each $\svs_i^n \in \calS_i^n$,
generate $\lceil 2^{nR_i} \rceil$  $\rvx_i^n$ sequence drawn according to
$\prod_{j=1}^{n}p(\svx_{ij}|\svs_{ij},\svq_j)$.
Call these sequences $\rvx_i^n(\rvq^n,\rvs_i^n,m_i)$
where $m_i \in \{1,2,\ldots,2^{nR_i}\}$, $i=1,2$. In this way, the codebooks are
generated at each encoder and revealed to the decoder.

Since the sequence $\rvq^n$ serves as time sharing sequence, it can be
assumed that the sequence $\rvq^n$ is known
at both the encoders and at the decoder without loss of generality.

\item \textbf{Encoding:} Encoder $i$, upon observing $\rvs_i^n$ at the output of
host source $i$ and time sharing random sequence
$\rvq^n$, sends message $\rvw_i \in \{1,2,\ldots,\lceil2^{nR_i}\rceil\}$
by transmitting the codeword
$\rvx_i^n(\rvq^n,\rvs_i^n,\rvw_i)$.
In this way, the codeword $\rvx_i^n$ is chosen and
transmitted from Encoder $i$ for
a given time sharing sequence
$\rvq^n$, a given host sequence $\rvs_i^n$, 
and a message $\rvw_i$.

\item \textbf{Decoding:} Fix $0 <\epsilon_1<\epsilon$. Since the decoder knows
the time sharing sequence $\rvq^n=\svq^n$,
the decoder, upon receiving the channel output $\rvy^n$, looks for a tuple
$(\rvx_1^n(\svq^n,\svs_1^n,m_1),\rvx_2^n(\svq^n,\svs_2^n,m_2))$  such that
$(\rvx_1^n(\svq^n,\svs_1^n,m_1),\rvx_2^n(\svq^n,\svs_2^n,m_2),\rvy^n)\in
T_{\epsilon}^n[\rvq,\rvs_1,\rvs_2,\rvx_1,\rvx_2,\rvy|\svq^n,\svs_1^n,\svs_2^n]$
 for all $(\svs_1^n,\svs_2^n)\in T_{\epsilon_1}^n[\rvs_1,\rvs_2]$.
If a unique vector of sequences exists, the decoder declares that
$(\hat{\rvw}_1,\hat{\rvw}_2,\hat{\rvs}_1^n,\hat{\rvs}_2^n)=(m_1,m_2, \svs_1^n,\svs_2^n)$.
Otherwise, the decoder declares an error.
In this way, the messages and the host sequences are decoded at the decoder.

\item \textbf{Probability of error:}
The average probability of error is given by the following
{\allowdisplaybreaks
\begin{align}
P_e^n & = \sum_{(\svs_1^n,\svs_2^n,\svq^n)\in \calS_1^n \times \calS_2^n \times
\calQ^n}p(\svq^n)p(\svs_1^n,\svs_2^n)\mathrm{Pr}[\mathrm{error}|(\svs_1^n,\svs_2^n,\svq^n)] \nonumber \\
& \leq \sum_{(\svq^n,\svs_1^n,\svs_2^n)\not\in
T_{\epsilon_1}^n[\rvq,\rvs_1,\rvs_2]}p(\svq^n)p(\svs_1^n,\svs_2^n) \nonumber \\
& +\sum_{(\svq^n,\svs_1^n,\svs_2^n)\in
T_{\epsilon_1}^n[\rvq,\rvs_1,\rvs_2]}p(\svs_1^n,\svs_2^n)p(\svq^n)\mathrm{Pr}[\mathrm{error}|
(\svs_1^n,\svs_2^n,\svq^n)]
\label{eqn:DRIE_pe_events}
\end{align}}

The first term, $\mathrm{Pr}[(\svq^n,\svs_1^n,\svs_2^n)\not\in T_{\epsilon_1}^n[\rvq,\rvs_1,\rvs_2]]$,
in the right hand side expression of (\ref{eqn:DRIE_pe_events}) goes to zero as
$n \rightarrow \infty$ by Lemma~\ref{thm:ST_lemma1}.

Without loss of generality, it can be assumed that the time-sharing sequence is $\svq^n$,
the output of the host source $i$ is $\tilde{\svs}_i^n$, and $\rvw_i =1$
is being transmitted from Encoder $i$.
Hence, the codeword $\rvx_i^n(\svq^n,\tilde{\svs}_i^n,1)$ is transmitted from Encoder $i$.
It is also assumed that the time-sharing random sequence $\rvq^n=\svq^n$
is known at both the encoders
and the decoder. Let $F$ be the event that $(\tilde{s}_1^n,\tilde{s}_2^n)$ and $\svq^n$ are
the output of the host source pair and time sharing sequence, respectively and
$(\svq^n,\svs_1^n,\svs_2^n)\in T_{\epsilon_1}^n[\rvq,\rvs_1,\rvs_2]$.

The following error events are considered  to compute
$\mathrm{Pr}[\mathrm{error}|F]$ and can be made to
approach zero as $n \rightarrow \infty$.
\begin{enumerate}
\item $E_1$: $(\rvx_1^n(\svq^n,\tilde{\svs}_1^n,1),\rvx_2^n(\svq^n,\tilde{\svs}_2^n,1),\rvy^n) \not \in $
$T_{\epsilon}^n[\rvq,\rvs_1,\rvs_2,\rvx_1,\rvx_2,\rvy|\svq^n,\tilde{\svs}_1^n,\tilde{\svs}_2^n]$ under the event
$F$. By using Lemma~\ref{thm:ST_lemma1},
we can show that $\mathrm{Pr}[E_{1}|F] \rightarrow 0$ as $n \rightarrow \infty$.
\item $E_2$:$(\rvx_1^n(q^n,\tilde{\svs}_1^n,m_1),\rvx_2^n(\svq^n,\tilde{\svs}_2^n,1),\rvy^n) \in
T_{\epsilon}^n[\rvq,\rvs_1,\rvs_2,\rvx_1,\rvx_2,\rvy|\svq^n,\tilde{\svs}_1^n,\tilde{\svs}_2^n]$ under the
event $F$ for all $m_1 \neq 1$.
It can be shown that $\mathrm{Pr}(E_{2}|F) \rightarrow 0$ as
$n \rightarrow \infty$ by using Lemma~\ref{thm:ST_lemma1} and
Lemma~\ref{thm:ST_lemma2} if $0 \leq R_1 < \mathbb{I}(\rvx_1;\rvy|\rvs_1,\rvs_2,\rvx_2,\rvq)$.
\item $E_3$:$(\rvx_1^n(\svq^n,\svs_1^n,m_1),\rvx_2^n(\svq^n,\tilde{\svs}_2^n,1),\rvy^n) \in
T_{\epsilon}^n[\rvq,\rvs_1,\rvs_2,\rvx_1,\rvx_2,\rvy|\svq^n,\svs_1^n,\tilde{\svs}_2^n]$ under the event
$F$ for all $m_1 \in M_1$ and for all $\svs_1^n \neq \tilde{\svs}_1^n$ and $\svs_1^n \in
T_{\epsilon_1}^n[\rvs_1,\rvs_2|\tilde{\svs}_2^n]$.
It can be shown that $\mathrm{Pr}(E_{3}|F) \rightarrow 0$ as
$n \rightarrow \infty$ by using Lemma~\ref{thm:ST_lemma1} and  Lemma~\ref{thm:ST_lemma2} if
$0 \leq R_1 < \mathbb{I}(\rvs_1,\rvx_1;\rvy|\rvs_2,\rvx_2,\rvq)-\mathbb{H}(\rvs_1|\rvs_2)$.
\item $E_4$ : $(\rvx_1^n(\svq^n,\tilde{\svs}_1^n,1),\rvx_2^n(\svq^n,\tilde{\svs}_2^n,m_2),\rvy^n) \in
T_{\epsilon}^n[\rvq,\rvs_1,\rvs_2,\rvx_1,\rvx_2,\rvy|\svq^n,\tilde{\svs}_1^n,\tilde{\svs}_2^n]$ under
the event $F$ for all $m_2 \neq 1$.
It can be shown that $\mathrm{Pr}(E_{4}|F) \rightarrow 0$ as $n \rightarrow \infty$
by using Lemma~\ref{thm:ST_lemma1} and  Lemma~\ref{thm:ST_lemma2}
if $0 \leq R_2 < \mathbb{I}(\rvx_2;\rvy|\rvs_1,\rvx_1,\rvs_2,\rvq)$.
\item $E_5$ :$(\rvx_1^n(\svq^n,\tilde{\svs}_1^n,1),\rvx_2^n(\svq^n,\svs_2^n,m_2),\rvy^n) \in
T_{\epsilon}^n[\rvq,\rvs_1,\rvs_2,\rvx_1,\rvx_2,\rvy|\svq^n,\tilde{\svs}_1^n,\svs_2^n]$  under the event
$F$ for all $m_2 \in M_2$, $\svs_2^n \neq \tilde{s}_2^n$, and
$\svs_2^n \in T_{\epsilon_1}^n[\rvs_1,\rvs_2|\tilde{\svs}_1^n]$.
It can be shown that $\mathrm{Pr}(E_{5}|F) \rightarrow 0$ as $n \rightarrow \infty$
by using Lemma~\ref{thm:ST_lemma1} and  Lemma~\ref{thm:ST_lemma2} if $0 \leq
R_2 < \mathbb{I}(\rvx_2,\rvs_2;\rvy|\rvs_1,\rvx_1,\rvs_2,\rvq)-\mathbb{H}(\rvs_2|\rvs_1)$.
\item $E_6$ :$(\rvx_1^n(\svq^n,\tilde{\svs}_1^n,m_1),\rvx_2^n(\svq^n,\svs_2^n,m_2),\rvy^n) \in
T_{\epsilon}^n[\rvq,\rvs_1,\rvs_2,\rvx_1,\rvx_2,\rvy|\svq^n,\tilde{\svs}_1^n,\svs_2^n]$
under the event $F$ for all $m_1\in M_1$,
$m_2 \in M_2$, $\svs_2^n \neq \tilde{\svs}_2^n$ and $\svs_2^n \in
T_{\epsilon_1}^n[\rvs_1,\rvs_2|\tilde{\svs}_1^n]$.
It can be shown that $\mathrm{Pr}(E_{6}|F) \rightarrow 0$ as $n \rightarrow \infty$
by using Lemma~\ref{thm:ST_lemma1} and  Lemma~\ref{thm:ST_lemma2}
if $R_1+R_2 < \mathbb{I}(\rvx_1,\rvs_2,\rvx_2;\rvy|\rvs_1,\rvq)-\mathbb{H}(\rvs_2|\rvs_1)$.
\item $E_7$ :$(\rvx_1^n(\svq^n,\svs_1^n,m_1),\rvx_2^n(\svq^n,\svs_2^n,m_2),\rvy^n) \in
T_{\epsilon}^n[\rvq,\rvs_1,\rvs_2,\rvx_1,\rvx_2,\rvy|\svq^n,\tilde{\svs}_1^n,\svs_2^n]$
under the event $F$ for all $m_1\in M_1$,
$m_2 \in M_2$, $(\svs_1^n,\svs_2^n) \neq (\tilde{\svs}_1^n,\tilde{\svs}_2^n)$,
and $(\svs_1^n,\svs_2^n) \in T_{\epsilon_1}^n[\rvs_1,\rvs_2]$.
It can be shown that $\mathrm{Pr}(E_{7}|F) \rightarrow 0$ as $n \rightarrow \infty$
by using Lemma~\ref{thm:ST_lemma1} and  Lemma~\ref{thm:ST_lemma2}
if $0 \leq R_1+R_2 < \mathbb{I}(\rvs_1,\rvx_1,\rvs_2,\rvx_2;\rvy|\rvq)-\mathbb{H}(\rvs_1,\rvs_2)$.
\item $E_{8}$ :$(\rvx_1^n(\svq^n,\svs_1^n,m_1),\rvx_2^n(\svq^n,\tilde{\svs}_2^n,m_2),\rvy^n) \in
T_{\epsilon}^n[\rvq,\rvs_1,\rvx_1,\rvs_2,\rvx_2,\rvy|\svq^n,\svs_1^n,\tilde{\svs}_2^n]$
under the event $F$ for all $m_1\neq 1$,
$m_2 \in M_2$, $\svs_1^n \neq \tilde{\svs}_1^n$, and $\svs_1^n \in
T_{\epsilon_1}^n[\rvs_1,\rvs_2|\tilde{\svs}_2^n]$.
It can be shown that $\mathrm{Pr}(E_{8}|F) \rightarrow 0$ as $n \rightarrow \infty$
by using Lemma~\ref{thm:ST_lemma1} and  Lemma~\ref{thm:ST_lemma2} if
$0\leq R_1+R_2 < \mathbb{I}(\rvs_1,\rvx_1,\rvx_2;\rvy|\rvs_2,\rvq)-\mathbb{H}(\rvs_1|\rvs_2)$.
\item $E_{9}$ :$(\rvx_1^n(\svq^n,\tilde{\svs}_1^n,m_1),\rvx_2^n(\svq^n,\tilde{\svs}_2^n,m_2),\rvy^n) \in
T_{\epsilon}^n[\rvq,\rvs_1,\rvs_2,\rvx_1,\rvx_2,\rvy|\svq^n,\tilde{\svs}_1^n,\tilde{\svs}_2^n]$
 under the event $F$ for all $m_1\neq 1$, and $m_2 \neq M_2$.
 It can be shown that $\mathrm{Pr}(E_{9}|F) \rightarrow 0$ as $n \rightarrow \infty$
 by using Lemma~\ref{thm:ST_lemma1} and  Lemma~\ref{thm:ST_lemma2}
 if $0\leq R_1+R_2 < \mathbb{I}(\rvx_1,\rvx_2;\rvy|\rvs_1,\rvs_2,\rvq)$.
\end{enumerate}

Then by using the union bound, $\mathrm{Pr}[\mathrm{error}|F] \leq \sum_{j=1}^{9}\mathrm{Pr}[E_j|F]$.
$\mathrm{Pr}[\mathrm{error}|F]$ goes to zero as $n \rightarrow \infty$ since
$\mathrm{Pr}(E_j) \rightarrow 0$, where $j =1$ to $9$, as $n \rightarrow \infty$
if rate pair $(R_1,R_2)$ satisfies (\ref{eqn:innerbound_distributedRIE}).
It can be concluded that $P_e^n \rightarrow 0$ as $n \rightarrow 0$
if rate pair $(R_1,R_2)$ satisfies (\ref{eqn:innerbound_distributedRIE}).
\item \textbf{Average distortions:}
We consider two cases in calculating the average distortion between
the host sequence $\rvs_i^n$ and the codeword $\rvx_i^n$ for any given message
$m_i$ and $\svq^n \in T_\epsilon^n[\rvq]$.
If $\rvx_i^n(\svq^n,\rvs_i^n,m_i)) \in T_{\epsilon}^n(\rvx_i|\svq^n,\rvs_i^n)$ for any
$(\svq^n,\rvs_1^n,\rvs_2^n) \in T_{\epsilon_1}^n[\rvq,\rvs_1,\rvs_2]$, then
the distortion between $\rvs_i^n$ and $\rvx_i^n$ is given by
{\allowdisplaybreaks
\begin{align}
d_i(\rvs_i^n, \rvx_i^n) &=     \frac{1}{n}\sum_{\svx_i,\svs_i} N(\svx_i,\svs_i|\rvs_i^n,
\rvx_i^n)d_i(\svs_i,\svx_i), \nonumber \\
                       &\leq   \sum_{\svx_i,\svs_i}p(\svs_i,\svx_i)d_i(\svs_i,\svx_i) +
		       \epsilon d_{i,max} \nonumber \\
		      &\leq    \Delta + \epsilon d_{i,\mathrm{max}}
\end{align}}
where $d_{i,\mathrm{max}}$ is the maximum distortion over the set $\calS_i \times \calX_i$.
If $\rvx_i^n(\svq^n,\rvs_i^n,m_i)) \in T_{\epsilon}^n(\rvx_i|\svq^n,\svs_i^n)$
for any $(\svq^n,\rvs_1^n,\rvs_2^n)
\in T_{\epsilon_1}^n[\rvq,\rvs_1,\rvs_2]$,
the distortion $d_i(\rvs_i^n, \rvx_i^n)$ can be upper bounded by $d_{i,max}$.
From error event $E_1$ given $F$, we can show that $\mathrm{Pr}[\rvx_i^n(\svq^n,\rvs_i^n,m_i))
\in T_{\epsilon}^n(\rvx_i|\svq^n,\rvs_i^n)]$
goes to zero as $n \rightarrow \infty$. We can then conclude that $\lim_{n \rightarrow \infty}
\mathbb{E}d_i(\rvs_i^n,f^n(\rvs_i^n,\rvw_i)) \leq \Delta_i$
by letting $\epsilon \rightarrow 0$ and $n \rightarrow \infty$.

This concludes that $\calR_{\mathrm{MAC,C}}^{\mathrm{i}}(\Delta_1,\Delta_2) \subseteq C_{\mathrm{MAC,C}}(\Delta_1,\Delta_2)$.
\end{itemize}
\subsection{Proof of Theorem  \ref{thm:outerbound_distributedRIE}}
\label{sec:proof_outerbound_DRIE}
We prove the following lemmas which will be used in the
proof of Theorem~\ref{thm:outerbound_distributedRIE}.
\begin{lemma}\label{lemma:outerbound_distributedRIE1}
Let $(\rvq_j,\rvs_1,\rvs_2,(\rvx_{1j},\rvx_{1j}),(\rvx_{2j},\rvx_{2j}),\rvy_j) \in
\calP_{\mathrm{MAC}}^i(\Delta_{1j},\Delta_{2j})$,  let
$\sum_{j=1}^n\lambda_j = 1$,  $\lambda_j >0$ for $j \in \{1,2,\ldots,n\}$, and
let $\Delta_i=\sum_{j=1}^{n}\lambda_j\Delta_{ij}$ for $i \in \{1,2\}$.
Then, there exists $$(\rvq,\rvs_1,\rvs_2,(\rvx_1,\rvx_1),(\rvx_2,\rvx_2),\rvy)
\in \calP_{\mathrm{MAC}}^i(\Delta_1,\Delta_2)$$ such that
\begin{subequations}\label{eqn:outerbound_DRIE_1}
\begin{align}
\sum_{j=1}^n\lambda_j[\mathbb{I}(\rvs_{1},\rvx_{1j};\rvy_j|\rvx_{2j},\rvs_{2},\rvq_j)]&=
\mathbb{I}(\rvs_1,\rvx_1;\rvy|\rvx_2,\rvs_2,\rvq) \\
\sum_{j=1}^n\lambda_j[\mathbb{I}(\rvs_2,\rvx_{2j};\rvy_j|\rvs_1,\rvx_{1j},\rvq_j)]&=
\mathbb{I}(\rvs_2,\rvx_2;\rvy|\rvx_1,\rvs_1,\rvq) \\
\sum_{j=1}^n\lambda_j[\mathbb{I}(\rvs_{1},\rvx_{1j},\rvs_{2},\rvx_{2j};\rvy_j|\rvq_j)] &=
\mathbb{I}(\rvs_1,\rvx_1,\rvs_2,\rvx_2;\rvy|\rvq)
\end{align}
\end{subequations}
\end{lemma}

\textbf{Proof:} If we prove the lemma for $n=2$, then
we can easily extend it to any value of $n$.
Let $n=2$ and let $\lambda_1+\lambda_2 =1$, $\lambda_j >0$ for $j=1,2.$
Let $\beta$ be a binary random variable such that
$\mathrm{Pr}(\rvz=j)=\lambda_j$ for $j=1,2.$
Let \[(\rvq,\rvs_1,\rvs_2,(\rvx_1,\rvx_1),(\rvx_2,\rvx_2),\rvy) =
((\rvz,\rvq_z),\rvs_1,\rvs_2,(\rvx_{1z},\rvx_{1z}),(\rvx_{2z},\rvx_{2z}),\rvy_z).\]
 \[ (\rvq,\rvs_1,\rvs_2,(\rvx_1,\rvx_1), (\rvx_2,\rvx_2),\rvy)=
 \begin{cases} ((\rvq_1,1),\rvs_1,\rvs_2,(\rvx_{11},\rvx_{11}),(\rvx_{21},\rvx_{21}),\rvy_1),
 &\text{if $\rvz =1$;}\\ ((\rvq_2,2),\rvs_1,\rvs_2, (\rvx_{12}, \rvx_{12}), 
 (\rvx_{22},\rvx_{22}), \rvy_2)&\text{if $\rvz=2$;}\\
 \end{cases} \]
To show that $(\rvq,\rvs_1,\rvs_2, (\rvx_1,\rvx_1), (\rvx_2,\rvx_2),\rvy)\in
\calP_{\mathrm{MAC}}^i(\Delta_{1},\Delta_{2})$,
we have to check the conditions in Definition (\ref{def:ib_pdfconstraints}).
We can easily show that
$(\rvq,\rvs_1,\rvs_2, (\rvx_1,\rvx_1),(\rvx_2,\rvx_2),\rvy)$ satisfies
the first condition. To check the second condition,
we observe that the $\rvx_1 \leftrightarrow (\rvs_1,\rvs_2,\rvq)
\leftrightarrow \rvx_2$ follows as consequence of
\[\mathbb{I}(\rvx_1,\rvx_2|\rvs_1,\rvs_2,\rvq)=
\lambda_1 \mathbb{I}(\rvx_{11},\rvx_{21}|\rvs_1,\rvs_2,\rvq_1)+
\lambda_2 \mathbb{I}(\rvx_{12},\rvx_{22}|\rvs_1,\rvs_2,\rvq_2) =0 \]
Similarly, $\rvx_1 \leftrightarrow (\rvs_1,\rvq) \leftrightarrow \rvs_2$ and
$\rvs_1 \leftrightarrow (\rvs_2,\rvq)
\leftrightarrow \rvx_2$.
We can easily verify that
$\mathbb{E}d_i(\rvs_i,\rvx_i)<\lambda_1\Delta_{i1} +\lambda_2\Delta_{i2}$,
for $i=1,2$ using the distribution on $(\rvq,\rvs_1,\rvs_2,(\rvx_1,\rvx_1),(\rvx_2,\rvx_2),\rvy)$.
Since the distribution on $(\rvq,\rvs_1,\rvs_2,(\rvx_1,\rvx_1),(\rvx_2,\rvx_2),\rvy)$
 satisfies the conditions in
Definition (\ref{def:ib_pdfconstraints}),
we can conclude that $(\rvq,\rvs_1,\rvs_2,(\rvx_1,\rvx_1),(\rvx_2,\rvx_2),\rvy)\in
\calP_{MAC}^i(\Delta_1,\Delta_2)$.
We can easily derive the equations (\ref{eqn:outerbound_DRIE_1})
by using the distribution on
$(\rvq,\rvs_1,\rvs_2,(\rvx_1,\rvx_1),(\rvx_2,\rvx_2),\rvy)$.
This completes the proof of Lemma.

\begin{lemma} \label{lem:outerbound_distributedRIE2}
Let $(\rvq_j,\rvs_1,\rvs_2,\rvx_{1j},\rvx_{2j},\rvy_j) \in
\calP_{\mathrm{MAC}}^o(\Delta_{1j},\Delta_{2j})$,  let
$\sum_{j=1}^n\lambda_j = 1$,  $\lambda_j >0$ for $j \in \{1,2,\ldots,n\}$, and
let $\Delta_i=\sum_{j=1}^{n}\lambda_j\Delta_{ij}$ for $i \in \{1,2\}$.
Then, there exists $(\rvq,\rvs_1,\rvs_2,\rvx_1,\rvx_2,\rvy)\in
\calP_{\mathrm{MAC}}^o(\Delta_1,\Delta_2)$ such that
{\allowdisplaybreaks
\begin{subequations}\label{eqn:outerbound_DRIE2}
\begin{align}
\sum_{j=1}^n\lambda_j[\mathbb{I}(\rvx_{1j},\rvs_{1};\rvy_j|\rvx_{2j},\rvs_{2},\rvq_j)]
&=\mathbb{I}(\rvx_1,\rvs_1;\rvy|\rvx_2,\rvs_2,\rvq)  \\
\sum_{j=1}^n\lambda_j[\mathbb{I}(\rvx_{2j},\rvs_{2j};\rvy_j|\rvx_{1j},\rvs_{1j},\rvq_j)]
&=\mathbb{I}(\rvx_2,\rvs_2;\rvy|\rvx_1,\rvs_1,\rvq)   \\
\sum_{j=1}^n\lambda_j[\mathbb{I}(\rvx_{1j},\rvs_{1j},\rvx_{2j},\rvs_{2j};\rvy_j|\rvq_j)]
&=[\mathbb{I}(\rvx_1,\rvs_1,\rvx_2,\rvs_2;\rvy|\rvq)]
\end{align}
\end{subequations}}
\end{lemma}
\textbf{Proof:} We do not prove the lemma because proof is similar to the proof of
Lemma~\ref{lemma:outerbound_distributedRIE1}.

\begin{lemma}\label{lem:outerbound_distributedRIE3}
$\calR_{\mathrm{MAC,C}}^{\mathrm{i}}(\Delta_1,\Delta_2) \subseteq
\calR^{\mathrm{i}}_{\mathrm{MAC,C}}(\Delta'_1,\Delta'_2)$  and
$\calR^{\mathrm{o}}_{\mathrm{MAC,C}}(\Delta_1,\Delta_2) \subseteq
\calR^{\mathrm{o}}_{\mathrm{MAC,C}}(\Delta'_1,\Delta'_2)$
for any $\Delta_1 \leq \Delta'_1$ and $\Delta'_2 \leq \Delta'_2$.
\end{lemma}
\textbf{Proof:} This lemma can be directly proved from the fact that
$\calP^i_{\mathrm{MAC}}(\Delta_1,\Delta_2)\subseteq
\calP^i_{\mathrm{MAC}}(\Delta'_1,\Delta'_2)$ and
$\calP^o_{\mathrm{MAC}}(\Delta_1,\Delta_2)
\subseteq \calP^o_{\mathrm{MAC}}(\Delta'_1,\Delta'_2)$.

We  are now ready to prove the Theorem \ref{thm:outerbound_distributedRIE},
i.e., prove that for any  sequence of MAC IE codes
$(\lceil2^{nR_1}\rceil, \lceil2^{nR_2}\rceil, D_1^{(n)}, D_2^{(n)}, n)$
with $\lim_{n \rightarrow \infty} P_e^n=0$ and
$\lim_{n \rightarrow \infty} D_i^{(n)} \leq \Delta_i$,
for $i=1,2$, the rates must satisfy (\ref{eqn:distributedRIE_capacity}).

Consider a given code of block length $n$.
The joint distribution on $\calW_1 \times \calW_2\times\calS_1^n \times \calS_2^n \times
 \calX_1^n \times \calX_2^n \times \calY^n$ is given by
 \begin{eqnarray*}
 \lefteqn{p(\svw_1,\svw_2,\svs_1^n,\svs_2^n,\svx_1^n,\svx_2^n,\svy^n) =}\\
 & &\frac{1}{2^{nR_1}}\frac{1}{2^{nR_2}}\left (\prod_{j=1}^{n}p(\svs_{1j},\svs_{2j})\right
)p(\svx_1^n|\svw_1,\svs_1^n)p(\svx_2^n|\svw_2,\svs_2^n)
\prod_{i=1}^{n}p(\svy_j|\svx_{1j},\svx_{2j},\svs_{1j},\svs_{2j}),
 \end{eqnarray*}
 where, $p(\svx_i^n|\svw_i,\svs_i^n)$ is $1$ if $\svx_i^n=f_i^n(\svw_i,\svs_i^n)$ and $0$  otherwise,
  for $i=1,2$. By Fano's inequality \cite{cover91:book},
  the conditional entropy of
 $(\rvw_1,\rvw_2,\rvs_1^n,\rvs_2^n)$ given $\rvy^n$ is bounded as
 \begin{equation}
 \mathbb{H}(\rvw_1,\rvw_2,\rvs_1^n,\rvs_2^n|\rvy^n)  \leq n(R_1+R_2+ \log_{2}(|\calS_1||\calS_2|))P_e^n+1
 \stackrel{\triangle}{=} n\epsilon_n,
 \end{equation}
for $i=1,2$, where $\epsilon_n \rightarrow 0$ as $P_e^n \rightarrow 0.$  We can now bound the rate $R_1$ as
{\allowdisplaybreaks
  \begin{align}
  nR_1 & \leq \mathbb{H}(\rvw_1)~= \mathbb{H}(\rvw_1|\rvw_2) \nonumber \\
       & \stackrel{(a)}{=} \mathbb{H}(\rvw_1,\rvs_1^n|\rvw_2,\rvs_2^n)
       -\mathbb{H}(\rvs_1^n|\rvs_2^n) \nonumber\\
       & = \mathbb{H}(\rvw_1,\rvs_1^n|\rvw_2,\rvs_2^n)
       - \mathbb{H}(\rvw_1,\rvs_1^n|\rvw_2,\rvs_2^n,\rvy^n) \nonumber \\
       & +\mathbb{H}(\rvw_1,\rvs_1^n|\rvw_2,\rvs_2^n\rvy^n)
	  -\mathbb{H}(\rvs_1^n|\rvs_2^n) \nonumber\\
       &\stackrel{(b)}{\leq} \mathbb{H}(\rvw_1,\rvs_1^n|\rvw_2,\rvs_2^n)-
       \mathbb{H}(\rvw_1,\rvs_1^n|\rvw_2,\rvs_2^n,\rvy^n)
       -\mathbb{H}(\rvs_1^n|\rvs_2^n)+n\epsilon_{n} \nonumber\\
       &\stackrel{(c)}{=} \mathbb{H}(\rvw_1,\rvs_1^n|\rvw_2,\rvx_2^n,\rvs_2^n)-
        \mathbb{H}(\rvw_1,\rvs_1^n|\rvy^n,\rvw_2,\rvx_2^n,\rvs_2^n)-
	\mathbb{H}(\rvs_1^n|\rvs_2^n)+n\epsilon_{n} \nonumber\\
       & = \mathbb{I}(\rvw_1,\rvs_1^n;\rvy^n|\rvw_2,\rvx_2^n,\rvs_2^n)-
       \mathbb{H}(\rvs_1^n|\rvs_2^n)+n\epsilon_{n} \nonumber\\
       &=\mathbb{H}(\rvy^n|\rvw_2,\rvx_2^n,\rvs_2^n)-
       \mathbb{H}(\rvy^n|\rvw_2,\rvx_2^n,\rvs_2^n,\rvw_1,\rvs_1^n)
       -\mathbb{H}(\rvs_1^n|\rvs_2^n)+n\epsilon_{n} \nonumber\\
       &\stackrel{(d)}{=}\mathbb{H}(\rvy^n|\rvw_2,\rvx_2^n,\rvs_2^n)-
       \mathbb{H}(\rvy^n|\rvw_2,\rvx_2^n,\rvs_2^n,\rvw_1,\rvs_1^n,\rvx_1^n)-
       \mathbb{H}(\rvs_1^n|\rvs_2^n)+n\epsilon_{n} \nonumber \\
       & \stackrel{(e)}{=} \sum_{j=1}^n
       [\mathbb{H}(\rvy_j|\rvw_2,\rvx_2^n,\rvs_2^n,\rvy^{j-1})-
      \mathbb{H}(\rvy_j|\rvw_2,\rvx_2^n,\rvs_2^n,\rvw_1,\rvs_1^n,\rvx_1^n,\rvy^{j-1}) \nonumber \\
      & - \mathbb{H}(\rvs_{1j}|\rvs_2^n,\rvs_{1}^{j-1})]+n\epsilon_{n} \nonumber\\
       & \stackrel{(f)}{=} \sum_{j=1}^n
       [\mathbb{H}(\rvy_j|\rvw_2,\rvx_2^n,\rvs_2^n,\rvy^{j-1})-
       \mathbb{H}(\rvy_j|\rvx_{1j},\rvs_{1j},\rvx_{2j},\rvs_{2j})-
       \mathbb{H}(\rvs_{1j}|\rvs_{2j})]+n\epsilon_{n} \nonumber\\
       & \stackrel{(g)}{\leq} \sum_{j=1}^n [\mathbb{H}(\rvy_j|\rvx_{2j},\rvs_{2j})-
       \mathbb{H}(\rvy_j|\rvx_{1j},\rvs_{1j},\rvx_{2j},\rvs_{2j})-
       \mathbb{H}(\rvs_{1j}|\rvs_{2j})]+n\epsilon_{n} \nonumber\\
       & = \sum_{j=1}^n [\mathbb{I}(\rvx_{1j},\rvs_{1j};\rvy_j|\rvx_{2j},\rvs_{2j})
       -\mathbb{H}(\rvs_{1j}|\rvs_{2j})]+n\epsilon_{n},\nonumber
   \end{align} }
where: \newline
$(a)$ follows from the fact that $\rvw_1$ is independent of each other; and
$(\rvw_1,\rvw_2)$ is independent of $(\rvs_1^n,\rvs_2^n)$. \newline
$(b)$ follows from Fano's inequality, \newline
$(c)$ follows from the fact that $\rvx_2^n$ is a function of $(\rvw_1,\rvs_1^n)$, \newline
$(d)$ follows from the fact that $\rvx_1^n$ is a function of $(\rvw_1,\rvs_1^n)$, \newline
$(e)$ follows from the chain rule of mutual information and entropy, \newline
$(f)$ follows from the fact that $\rvy_j$ depends only on $\rvx_{1j}$,
$\rvx_{2j}$, $\rvs_{1j}$, and $\rvs_{2j}$ by the memoryless property of the channel and
$\rvs_{1j} \leftrightarrow \rvs_{2j} \leftrightarrow (\rvs_{1}^{j-1},\rvs_{2}^{j-1},\rvs_{2,j+1}^n)$,
 \newline
$(g)$ follows from removing conditioning.

Hence, we have
\begin{align}
R_1 & \leq \frac{1}{n} \sum_{j=1}^n [\mathbb{I}(\rvx_{1j},\rvs_{1};\rvy_j|\rvx_{2j},\rvs_{2})]-
\mathbb{H}(\rvs_{1}|\rvs_{2})]+\epsilon_{n} \nonumber \end{align}

Similarly, we can bound $R_2$ and $R_1+R_2$ as
{\allowdisplaybreaks
\begin{subequations}
\begin{align}
R_2 & \leq \frac{1}{n} \sum_{j=1}^n [\mathbb{I}(\rvx_{2j},\rvs_{2};\rvy_j|\rvx_{1j},\rvs_{1})]-
\mathbb{H}(\rvs_{1}|\rvs_{2})+\epsilon_{n}, \nonumber \\
R_1+R_2 & \leq \frac{1}{n}\sum_{j=1}^n[\mathbb{I}(\rvx_{1j},\rvs_{1j},\rvx_{2j},\rvs_{2};\rvy_j)]-
\mathbb{H}(\rvs_{1}|\rvs_{2})+\epsilon_{n}.\nonumber
\end{align}
\end{subequations}}

If the host random variables $\rvs_1$ and $\rvs_2$ are correlated,  we can clearly see that
the random vector $(\rvq_{j},\rvs_1,\rvs_2,\rvx_{1j},\rvx_{2j},\rvy_j)$ 
with $p(\svq_j=j)=1$ belongs to set \\
$\calP^o_{\mathrm{MAC}}(\mathbb{E}[d_1(\rvs_{1j},\rvx_{1j})],\mathbb{E}[d_2(\rvs_{2j},\rvx_{1j}]))$ for $j\in\{1,2,\dots,n\}$.
According to Lemma~\ref{lem:outerbound_distributedRIE2}, there exists a random vector
$(\rvq,\rvs_1,\rvs_2,\tilde{\rvx}_1, \tilde{\rvx}_2,\tilde{\rvy})
\in \calP^o_{\mathrm{MAC}}(\frac{1}{n}\sum_{j=1}^n \mathbb{E}[d_1(\rvs_{1j},\rvx_{1j})],\frac{1}{n}
\sum_{j=1}^n \mathbb{E}[d_2(\rvs_{1j},\rvx_{1j})])$ such that the following is true
{\allowdisplaybreaks
\begin{subequations} \label{eqn:outerbound_DRIE3}
\begin{align}
\frac{1}{n}\sum_{j=1}^n [\mathbb{I}(\rvx_{1j},\rvs_{1};\rvy_j|\rvx_{2j},\rvs_{2})] & =
\mathbb{I}(\tilde{\rvx}_{1},\rvs_{1};\tilde{\rvy}|\tilde{\rvx}_{2},\rvs_{2},\rvq) \nonumber \\
\frac{1}{n}\sum_{j=1}^n [\mathbb{I}(\rvx_{2j},\rvs_{2};\rvy_j|\rvx_{1j},\rvs_{1})] & =
\mathbb{I}(\tilde{\rvx}_{2},\rvs_{2};\tilde{\rvy}|\tilde{\rvx}_{1},\rvs_{1},\rvq) \nonumber\\
\frac{1}{n}\sum_{j=1}^n[\mathbb{I}(\rvx_{1j},\rvs_{1j},\rvx_{2j},\rvs_{2};\rvy_j)] &=
\mathbb{I}(\tilde{\rvx}_{1},\rvs_{1},\tilde{\rvx}_{2},\rvs_{2};\tilde{\rvy}|\rvq) \nonumber
\end{align}
\end{subequations}}
As $n \rightarrow \infty$, we can conclude the following
\begin{align}
\calC_{\mathrm{MAC,C}}(\Delta_1,\Delta_2) & \subseteq \calR_{\mathrm{MAC,C}}^{\mathrm{o}}\left(\lim_{n \rightarrow
\infty}\frac{1}{n}\sum_{j=1}^n \mathbb{E}[d_1(\rvs_{1j},\rvx_{1j})],\lim_{n \rightarrow
\infty}\frac{1}{n}\sum_{j=1}^n \mathbb{E}[d_2(\rvs_{1j},\rvx_{1j})]\right) \nonumber \\
& \stackrel{(a)}{\subseteq}\calR_{\mathrm{MAC,C}}^{\mathrm{o}}(\Delta_1,\Delta_2)
\end{align}
where $(a)$ follows from the Lemma~\ref{lem:outerbound_distributedRIE3}.

If the host random variables $\rvs_1$ and $\rvs_1$ are independent,
we can obtain the following from the condition that the messages $\rvw_1$ and $\rvw_2$ are independent.
\[p(\svx_{1j},\svx_{2j}|\svs_{1j},\svs_{2j})=p(\svx_{1j}|\svs_{1j})p(\svx_{2j}|\svs_{2j}).\]
Then, we can clearly see that the random variable tuple
$(\rvq_{j},\rvs_1,\rvs_2,(\rvx_{1j},\rvx_{1j}),(\rvx_{2j},\rvx_{2j}),\rvy_j)$ with $p(\svq_j=j)=1$ belongs to set
$\calP_{\mathrm{MAC}}^i(\mathbb{E}[d_1(\rvs_{1j},\rvx_{1j})],\mathbb{E}[d_2(\rvs_{2j},\rvx_{1j})])$ for $j\in\{1,2,\dots,n\}$.
According to Lemma~\ref{lemma:outerbound_distributedRIE1}, there exists a random vector
$$(\rvq,\rvs_1,\rvs_2,(\tilde{\rvx}_1,\tilde{\rvx}_1), (\tilde{\rvx}_2,\tilde{\rvx}_2),\tilde{\rvy}) \in
\calP_{\mathrm{MAC}}^i(\frac{1}{n}\sum_{j=1}^n \mathbb{E}[d_1(\rvs_{1j},\rvx_{1j})],\frac{1}{n}
\sum_{j=1}^n \mathbb{E}[d_2(\rvs_{1j},\rvx_{1j})])$$
such that (\ref{eqn:outerbound_DRIE3}) is true.
As $n \rightarrow \infty$, we can conclude the following
\begin{align}
 \calC_{\mathrm{MAC,C}}(\Delta_1,\Delta_2) & \subseteq \calR^{\mathrm{i}}_{\mathrm{MAC,C}}\left(\lim_{n \rightarrow
\infty}\frac{1}{n}\sum_{j=1}^n \mathbb{E}[d_1(\rvs_{1j},\rvx_{1j})],\lim_{n \rightarrow
\infty}\frac{1}{n}\sum_{j=1}^n \mathbb{E}[d_2(\rvs_{1j},\rvx_{1j})\right)] \nonumber \\
& \stackrel{(a)}{\subseteq}\calR^{\mathrm{i}}_{\mathrm{MAC,C}}(\Delta_1,\Delta_2)
\end{align}
where $(a)$ follows from the Lemma~\ref{lem:outerbound_distributedRIE3}.
This completes the proof of Theorem~\ref{thm:outerbound_distributedRIE}.

\subsection{Proof of Theorem~\ref{thm:innerbound_schemeB}}
\label{sec:proof_innerbound_schemeB}
In this section, we show that $\calR_{B'}^i(\Delta)\subseteq
\calC_{B'}(\Delta)$. Fix the random vector $(\rvu,\rvs,\rvx,\rvy,\rvz)
\in \calP(\Delta)$. For each $n$, we construct a
$(\lceil2^{nR_1}\rceil,\lceil2^{nR_2}\rceil,D^{(n)},n)$ BC IE
code as follows.

\begin{itemize}
\item \textbf{Code construction :}
Generate $\lceil2^{nR_2}\rceil2^{n(\mathbb{I}(\rvu;\rvs)+\epsilon)}$ $\rvu^n$
sequences drawn according to $\prod_{j=1}^{n}p(\svu_{j})$.  Distribute
these sequences randomly into $\lceil2^{nR_2}\rceil$ bins such that
each bin has $2^{n(\mathbb{I}(\rvu;\rvs)+\epsilon)}$ sequences. Label all
sequences $\rvu_1^n$ in bin $m_2 \in
\{1,2,\ldots,\lceil2^{nR_2}\rceil\}$ as $\rvu_1^n(m_2)$. For each
$(\rvs^n,\rvu^n) \in T_{\epsilon}^n[\rvs,\rvu]$, generate $\lceil
2^{nR_1}\rceil$ $\rvx^n$ sequences according to
$\prod_{j=1}^{n}p(x_j|u_j,s_j)$. Label these sequences as
$\rvx^n(\rvs^n,\rvu^n,m_1)$, where $(\rvs^n,\rvu^n) \in
T_{\epsilon}^n[\rvs,\rvu]$ and $m_1 \in \{1,2,\ldots,\lceil
2^{nR_1}\rceil\}.$ These codebooks are revealed to the encoder and
both the decoders.

\item \textbf{Encoder :} The encoder, upon observing $\rvs^n \in
T_{\epsilon}^n[\rvs]$ at the output of the host source, embeds message
$\rvw_2 \in \{1,2,\ldots,\lceil2^{nR_2}\rceil\}$ into the host
sequence by looking for a $\rvu^n$ in bin $\rvw_2$ such that
$\rvu^n(\rvw_2) \in T_{\epsilon}^n[\rvs,\rvu|\rvs^n]$. If such a
sequence $\rvu^n(\rvw_2)$ does not exist, the encoder declares an
error; otherwise, the encoder embeds message $\rvw_1 \in
\{1,2,\ldots,\lceil2^{nR_1}\rceil\}$ into the host sequence $\rvs^n$
by choosing the codeword $\rvx^n(\rvs^n,\rvu^n(\rvw_2),\rvw_1)$.

\item \textbf{Decoder~1:} Decoder~1, upon receiving  $\rvy^n$, which
is a distorted or attacked version of the embedded sequence $\rvx^n$,
looks for $\rvu^n(m_2)$, $m_2 \in \{1,2,\ldots,\lceil2^{nR_2}\rceil\}
$ such that $(\rvu^n(m_2),\rvy^n) \in T_{\epsilon}^n[\rvu,\rvy]$. If a
unique codeword $\rvu^n(m_2)$ does not exist, Decoder~1 declares an
error; otherwise, Decoder~1 declares that $\hat{\rvw}_2=m_2$.  Upon
decoding the sequence $\rvu^n(\hat{\rvw}_2)$, Decoder~1 looks for
$\rvx^n(\svs^n,\rvu^n(\hat{\rvw}_2),m_1)$ such that
$(\rvx^n(\svs^n,\rvu^n(\hat{\rvw}_2),m_1),\rvy^n)\in
T_{\epsilon}^n[\rvs,\rvu, \rvx,\rvy|\svs^n,\rvu^n(\hat{\rvw}_2)]$ for
each $\svs^n \in T_{\epsilon}^n[\rvu,\rvs|\rvu^n(\hat{\rvw}_2)]$ and $m_1
\in \{1,2,\ldots,\lceil2^{nR_1}\rceil\}$. If a unique codeword
$\rvx^n(\svs^n,\rvu^n(\hat{\rvw}_2),m_1)$ exists, Decoder~1 declares that
$(\hat{\rvw}_1,\hat{\rvs}_2^n)=(m_1,\svs^n)$; otherwise, it declares an error.

\item \textbf{Decoder~2:}  Decoder~2, up on receiving  $\rvz^n$, which
is a degraded version of $\rvy^n$, looks for $\rvu^n(m_2)$, $m_2 \in
\{1,2,\ldots,\lceil2^{nR_2}\rceil\} $ such that
$(\rvu^n(m_2),\rvz^n)\in T_{\epsilon}^n[\rvu,\rvz]$. If a unique
codeword $\rvu^n(m_2)$ exists, Decoder~2 declares that
$\hat{\rvw}_2=m_2$; otherwise, Decoder~2 declares an error.

\item \textbf{Probability of error:} The average probability of error
is given by
{\allowdisplaybreaks
\begin{align}
P_e^n =& \sum_{\svs^n\in \calS^n }p(\svs^n)\mathrm{Pr}[\mathrm{error}|\svs^n] \nonumber \\
 \leq & \sum_{\svs^n\not\in T_{\epsilon}^n[\rvs]}p(\svs^n)+
 \sum_{\svs^n\in T_{\epsilon}^n[\rvs]}p(\svs^n)\mathrm{Pr}[\mathrm{error}|\svs^n],
  \label{eqn:CaseB_pe_events}
\end{align}}
where the first term, $\mathrm{Pr}[\svs^n\not\in T_{\epsilon}^n[\rvs]]$,
goes to zero as $n \rightarrow \infty$ by the strong asymptotic
equipartition property (AEP). Without loss of generality, it can be
assumed that the output of the host source is $\tilde{\svs}^n$, and the
message pair $(\rvw_1,\rvw_2)=(1,1)$ is to be embedded in to the host
sequence $\tilde{\svs}^n$. Let $F$ be the event that the host source
output is $\tilde{\svs}^n$. To compute $\mathrm{Pr}[\mathrm{error}|F]$,
let us write the error event as $E_0 \cup E_1 \cup E_2 \cup E_3$,
where:
\begin{enumerate}
\item $E_0$ is the event that there is no $\rvu^n(1)$ such that
$\rvu^n(1) \in T_{\epsilon}^n[\rvu,\rvs|\tilde{\svs}^n]$. Using
well-known rate-distortion arguments, the probability of this event
approaches zero as $n$ goes to infinity since each bin has
$2^{n(\mathbb{I}(\rvu;\rvs)+\epsilon)}$ $\rvu^n$ sequences.

Conditioned on the event $F \cap E_0^{c}$, it can also be assumed that
$\tilde{\rvu}^n(1)$ is jointly strongly typical with the host sequence
$\tilde{\svs}^n$. Hence, the embedded sequence
$\rvx^n(\tilde{s}^n,\tilde{\rvu}^n(1),1)$ is generated and transmitted
from the encoder.

\item $E_1$ is the event that
$$(\tilde{\rvu}^n(1),\rvx^n(\tilde{\svs}^n\!,\tilde{\rvu}^n(1),1),\rvy^n\!,\rvz^n)
\! \not \in \!  T_{\epsilon}^n[\rvs,\rvu,\rvx,\rvy,\rvz|\tilde{\svs}^n].$$
By the strong AEP, we can show that $\mathrm{Pr}[E_1|F\cap E_0^c]
\rightarrow 0$ as $n \rightarrow
\infty$.

\item $E_2 := E_{2,1} \cup  (E_{2,1}^c\cap E_{2,2})$, where
$E_{2,1}$ is the event that $(\rvu^n,\rvy^n) \in
T_{\epsilon}^n[\rvu,\rvy]$ for $\rvu^n \neq \tilde{\rvu}^n(1)$, and
$E_{2,2}$ is the event that
$(\rvx^n(\svs^n,\tilde{\rvu}^n(1),m_1),\rvy^n) \in
T_{\epsilon}^n[\rvs,\rvu,\rvx,\rvy|\rvs^n,\tilde{\rvu}^n(1)]$ for $m_1
\neq 1$ or $\svs^n \in \{ \svs^n:  \svs^n \neq \tilde{s}^n, \svs^n \in
T_{\epsilon}^n[\rvu,\rvs|\tilde{\rvu}^n(1)]\}$. It can be shown that
$\mathrm{Pr}[E_{2,1}|F\cap E_0^{c}] \rightarrow 0$ as $n \rightarrow
\infty$ if $R_2 \leq \mathbb{I}(\rvu;\rvy)-\mathbb{I}(\rvu;\rvs)$ and that
$\mathrm{Pr}(E_{2,2}|F\cap E_0^{c}\cap E_{2,1}^c)
\rightarrow 0$ as $n \rightarrow \infty$ if $R_1 \leq
\mathbb{I}(\rvs,\rvx;\rvy|\rvu)-\mathbb{H}(\rvs|\rvu)$.

\item $E_3$ is the event that $(\rvu^n,\rvz^n) \in T_{\epsilon}^n[\rvu,\rvz]$
for $\rvu^n \neq \tilde{\rvu}^n(1)$. Using Gel'fand-Pinsker arguments,
it can be shown that $\mathrm{Pr}[E_3|F\cap E_0^{c}] \rightarrow 0$ as
$n \rightarrow \infty$ if $R_2 \leq \mathbb{I}(\rvu;\rvz)-\mathbb{I}(\rvu;\rvs)$.
Because the broadcast channel is degraded, this constraint on $R_2$ is
more restrictive than the previous constraint.
\end{enumerate}

Thus, by the union bound, it can be shown that $P_e^n$ goes to zero as
$n \rightarrow \infty $ if $(R_1,R_2) \in \calR^i_{B'}$.

\item \textbf{Average distortion:} Since $(\rvx^n,\tilde{\svs}^n)$ is
jointly strongly typical with high probability and the distribution
belongs to $\calP(\Delta)$, it can be shown that the average
distortion $D^{(n)}$ associated with the generated code satisfies the
distortion constraint $\Delta$ as $n \rightarrow \infty$ as i
n the Proof of Theorem~\ref{thm:innerbound_distributedRIE}.

\end{itemize}
\subsection{Proof of Theorem~\ref{thm:outerbound_schemeB}}
\label{sec:proof_outerbound_schemeB}
In this section, we show that $\calC_{B'}(\Delta)\subseteq\calR_{B'}^o(\Delta)$.
If we are given a sequence of
$(\lceil2^{nR_1}\rceil,\lceil2^{nR_2}\rceil,D^{(n)},n)$ BC IE
codes, i.e., $\rvx^n=f(\rvw_1,\rvw_2,\rvs^n)$, $g_{1,B'}^n(\rvy^n)=(\hat{\rvw}_1,\hat{\rvw}_2,\hat{\rvs}^n)$,
 and $g_{2,B'}^n(\rvz^n)=\hat{\rvw}_2$, with $\lim_{n \rightarrow \infty} P_e^n=0$ and $\lim_{n
\rightarrow \infty} D^{(n)} \leq \Delta$, then we show that the rate pair $(R_1,R_2)$
must satisfy (\ref{eqn:outerbound_schemeB}) for some
$((\rvu,\rvv),\rvs,\rvx,\rvy,\rvz) \in \calP(\Delta)$. Consider a
given code of block length $n$. The joint distribution on $\calW_1
\times \calW_2\times\calS^n \times \calX^n \times \calY^n \times
\calZ^n$ induced by the code is given by
\begin{eqnarray*}
 \lefteqn{p(\svw_1,\svw_2,\svs^n,\svx^n,\svy^n,\svz^n) =}\\
 & &\frac{1}{\lceil2^{nR_1}\rceil \lceil 2^{nR_2}\rceil}p(\svs^n)
 p(\svx^n|\svw_1,\svw_2,\svs^n) \\
 & &\times \prod_{i=1}^{n}p(\svy_j|\svx_{j},\svs_{j})p(\svz_j|\svy_j),
 \end{eqnarray*}
 where, $p(\svx^n|\svw_1,\svw_2,\svs^n)$ is $1$ if $\svx^n=
 f^n(\svw_1,\svw_2,\svs^n)$ and $0$  otherwise.
 We can bound the rate $R_1$ as follows:
 {\allowdisplaybreaks
 \begin{align}
 nR_1 \leq & \mathbb{H}(\rvw_1) \nonumber \\
      \stackrel{(a)}{=} & \mathbb{H}(\rvw_1,\rvs^n|\rvw_2) -
      \mathbb{H}(\rvs^n|\rvw_2) \nonumber \\
      = & \mathbb{H}(\rvw_1,\rvs^n|\rvw_2) -
      \mathbb{H}(\rvw_1,\rvs^n|\rvw_2,\rvy^n) \nonumber \\
      &+ \mathbb{H}(\rvw_1,\rvs^n|\rvw_2,\rvy^n)-
      \mathbb{H}(\rvs^n|\rvw_2) \nonumber \\
     \stackrel{(b)}{\leq} & \mathbb{I}(\rvw_1,\rvs^n;\rvy^n|\rvw_2)-
     \mathbb{H}(\rvs^n|\rvw_2) +n\epsilon_n \nonumber \\
     \stackrel{(c)}{=}&\sum_{j=1}^n [\mathbb{I}(\rvw_1,\rvs^n;\rvy_j|\rvw_2,\rvy^{j-1})
      -\mathbb{H}(\rvs_j|\rvw_2)] +n\epsilon_n \nonumber \\
     \stackrel{(d)}{=}& \sum_{j=1}^n [\mathbb{H}(\rvy_j|\rvw_2,\rvy^{j-1})
     -\mathbb{H}(\rvy_j|\rvw_2,\rvy^{j-1},\rvw_1,\rvs^n,\rvx^n)\nonumber \\
     &\qquad -\mathbb{H}(\rvs_j|\rvw_2)] +n\epsilon_n \nonumber \\
     \stackrel{(e)}{=}& \sum_{j=1}^n   [\mathbb{H}(\rvy_j|\rvw_2,\rvy^{j-1},\rvz^{j-1}) -
     \mathbb{H}(\rvy_j|\rvs_j,\rvx_j) \nonumber \\
     & \qquad -\mathbb{H}(\rvs_j|\rvw_2)] +n\epsilon_n \nonumber \\
     \stackrel{(f)}{\leq}&\sum_{j=1}^n [ \mathbb{H}(\rvy_j|\rvw_2,\rvz^{j-1}) -
     \mathbb{H}(\rvy_j|\rvs_j,\rvx_j,\rvw_2,\rvz^{j-1}) \nonumber \\
     & \qquad-\mathbb{H}(\rvs_j|\rvw_2,\rvz^{j-1})] +n\epsilon_n \nonumber \\
     =&\sum_{j=1}^n  \mathbb{I}(\rvs_j,\rvx_j;\rvy_j|\rvw_2,\rvz^{j-1})
     -\mathbb{H}(\rvs_j|\rvw_2,\rvz^{j-1})
     +n\epsilon_n \label{eqn:temp1_outerbound_schemeB}
      \end{align}}
\noindent where, $\epsilon_n \rightarrow 0$ as $n \rightarrow \infty$, and\\
(a) follows from the fact that $\rvw_1$, $\rvw_2$ and $\rvs^n$ are mutually independent, \\
(b) follows from Fano's inequality, \\
(c) follows from the chain rule and the fact that $\rvs^n$ is i.i.d.\
and independent of $\rvw_2$, \\
(d) follows from the fact that $\rvx^n$ is a deterministic function of $(\rvw_1,\rvw_2,\rvs^n)$, \\
(e) follows from degraded and memoryless properties of the broadcast
channel, and \\
(f) follows from removing conditioning in the positive term and
introducing conditioning in the negative term.\\

 We can also bound the rate $R_2$ as follows:
 {\allowdisplaybreaks
 \begin{align}
 nR_2  \leq &\mathbb{H}(\rvw_2) \nonumber \\
      \stackrel{(a)}{\leq} & \mathbb{I}(\rvw_2;\rvz^n) + n\epsilon_n \nonumber \\
      =& \sum_{j=1}^{n} [\mathbb{I}(\rvw_2,\rvs_{j+1}^n;\rvz^{j})
      - \mathbb{I}(\rvw_2,\rvs_{j}^n;\rvz^{j-1})] + n\epsilon_n \nonumber \\
      \stackrel{(b)}{\leq}& \sum_{j=1}^{n} [\mathbb{I}(\rvw_2,\rvs_{j+1}^n;\rvz^{j-1})+
      \mathbb{I}(\rvw_2,\rvs_{j+1}^n;\rvz_j|\rvz^{j-1}) \nonumber \\
      &\quad\, -\mathbb{I}(\rvw_2,\rvs_{j+1}^n;\rvz^{j-1})
      -\mathbb{I}(\rvs_j;\rvz^{j-1}|\rvw_2,\rvs_{j+1}^n)] + n\epsilon_n  \nonumber \\
      =& \sum_{j=1}^{n} [\mathbb{I}(\rvw_2,\rvs_{j+1}^n;\rvz_j|\rvz^{j-1})
      -\mathbb{I}(\rvs_j;\rvz^{j-1}|\rvw_2,\rvs_{j+1}^n)] + n\epsilon_n  \nonumber \\
      =&\sum_{j=1}^{n} [\mathbb{H}(\rvz_j|\rvz^{j-1})
      -\mathbb{H}(\rvz_j|\rvw_2,\rvz^{j-1},\rvs_{j+1}^n)\nonumber\\
      &\quad\, -\mathbb{H}(\rvs_j|\rvw_2,\rvs_{j+1}^n)
      +\mathbb{H}(\rvs_j|\rvw_2,\rvz^{j-1},\rvs_{j+1}^n)]+ n\epsilon_n  \nonumber \\
       \stackrel{(c)}{\leq}& \sum_{j=1}^n [ \mathbb{H}(\rvz_j)
       - \mathbb{H}(\rvz_j|\rvw_2,\rvz^{j-1},\rvs_{j+1}^n) \nonumber\\
       &\quad\, -\mathbb{H}(\rvs_j) +\mathbb{H}(\rvs_j|\rvw_2,\rvz^{j-1},\rvs_{j+1}^n)]
        + n\epsilon_n \nonumber \\
       =& \sum_{j=1}^n[\mathbb{I}(\rvw_2,\rvz^{j-1},\rvs_{j+1}^n;\rvz_j)
       -\mathbb{I}(\rvw_2,\rvz^{j-1},\rvs_{j+1}^n;\rvs_j)] + n\epsilon_n
       \label{eqn:temp2_outerbound_schemeB}
\end{align}}
\noindent where, $\epsilon_n \rightarrow 0$ as $n \rightarrow \infty$, and \\
(a) follows from Fano's inequality, \\
(b) follows from applying the chain rule on $(\rvz^{j-1},\rvz_j)$ and
$(\rvs_{j+1}^n,\rvs_j)$ in the first and second mutual information expressions, respectively, and\\
(c) follows from removing conditioning and the fact that $\rvs^n$ is
i.i.d.\ and independent of $\rvw_2$.

Let $\tilde{\rvu}_j :=\{\rvw_2,\rvz^{j-1}\}$ and $\rvv_j
:=\{\rvs_{j+1}^n\}$ for $j=1,2,\ldots, n$. We can then write
(\ref{eqn:temp1_outerbound_schemeB}) and
(\ref{eqn:temp2_outerbound_schemeB}) as
{ \allowdisplaybreaks
\begin{subequations} \label{eqn:temp3_outerbound_schemeB}
\begin{align}
R_1 \leq & \mathbb{I}(\rvs,\rvx;\rvy|\rvq,\tilde{\rvu})
-\mathbb{H}(\rvs|\rvq,\tilde{\rvu})+ \epsilon_n,  \\
R_2 \leq & \mathbb{I}(\tilde{\rvu},\rvv;\rvz|\rvq)
-\mathbb{I}(\tilde{\rvu},\rvv;\rvs|\rvq)]+ \epsilon_n,
\end{align}
\end{subequations}}
where $\rvq$ takes values in the set $\calQ \in \{1,2,\dots,n\}$ with
equal probability and the joint probability distribution on
$(\rvs,\rvq,\tilde{\rvu},\rvv,\rvx,\rvy,\rvz)$ is
$p(\rvs=\svs,\rvq=\svq,\tilde{\rvu}=\tilde{\svu},\rvv=\svv,\rvx=\svx)
p(\svy|\svx,\svs)p(\svz|\svy),$
with
\begin{eqnarray*}
\lefteqn{p(\rvs=\svs,\rvq=\svq,\tilde{\rvu}=\tilde{\svu},\rvv=\svv,\rvx=\svx)=}\\
& &p(\svs)p(\svq)p(\rvu_q=\tilde{\svu},\rvv_q=\svv|\svs,\svq)p(\rvx_q=\svx|\svs,\svq,\tilde{\svu},\svv).
\end{eqnarray*}

Finally, we can write (\ref{eqn:temp3_outerbound_schemeB}) as
{\allowdisplaybreaks
\begin{align}
R_1 \leq & \mathbb{I}(\rvs,\rvx;\rvy|\rvu)
-\mathbb{H}(\rvs|\rvu)+ n\epsilon_n, \nonumber \\
R_2 \leq & \mathbb{I}(\rvu,\rvv;\rvz)
-\mathbb{I}(\rvu,\rvv;\rvs)+ n\epsilon_n, \nonumber
\end{align}}
where $\rvu:=(\rvq,\tilde{\rvu})$, since $\mathbb{I}(\tilde{\rvu},\rvv;\rvz|\rvq)
 \leq \mathbb{I}(\rvq,\tilde{\rvu},\rvv;\rvz)$ and $\mathbb{I}(\rvq;\rvs) = 0$.

Given any $\delta >0$, the associated distortion $D^{(n)}$, for
sufficiently large $n$, satisfies
{\allowdisplaybreaks
\begin{align}
\Delta +\delta \geq & D^{(n)}\nonumber \\
                  =   & \mathbb{E}d(\rvx^n,\rvs^n) \nonumber \\
                  =   & \frac{1}{n} \sum_{j=1}^n \sum_{\svx,\svs} p(\rvx_j=\svx,\rvs_j=\svs)
		  d(\svx,\svs) \nonumber \\
		  =  & \sum_{\svx,\svs} p(\rvx=\svx,\rvs=\svs) d(\svx,\svs) \nonumber \\
		  = & \mathbb{E}d(\rvx,\rvs). \nonumber
\end{align}}
As $n \rightarrow \infty$ and $\delta \rightarrow 0$,
$((\rvu,\rvv),\rvs,\rvx,\rvy,\rvz) \in \calP(\Delta)$ and $(R_1,R_2)
\in \calR_{B'}^o$. Thus, $\calC_{B'}(\Delta) \subseteq \calR^o_{B'}$.
\subsection{Proof of Theorem  \ref{thm:capacity_schemeC}}
\label{sec:proof_capacity_schemeC}
\subsubsection{Achievability}
In this section, we show that $\calR_{C'}^i(\Delta)\subseteq
\calC_{C'}(\Delta)$. Fix the random vector $(\rvu,\rvs,\rvx,\rvy,\rvz)
\in \calP(\Delta)$. For each $n$, we construct a
$(\lceil2^{nR_1}\rceil,\lceil2^{nR_2}\rceil,D^{(n)},n)$ BC IE
code as follows.
\begin{itemize}
\item \textbf{Code construction:}
At Encoder, for each $\svs^n \in \calS^n$, generate $2^{nR_2}$
$\rvu^n$ sequences drawn according to
$\prod_{j=1}^{n}p(\svu_{j}|\svs_{j})$. Denote these sequences as
$\rvu^n(\svs^n,m_2)$, where $m_2 \in \{1,2,\ldots,2^{nR_2}\}$
For each pair $(\svs^n,\rvu^n)$,
generate $2^{nR_1}$ $\rvx_1^n$ sequences drawn according to
$\prod_{j=1}^{n}p(\svx_{j}|\svu_j,\svs_{j})$.
Call these sequences $\rvx^n(\rvs^n,m_1,m_2)$ where
$m_1 \in \{1,2,\ldots,2^{nR_1}\}$. In this way, the codebook is
generated at the encoder and revealed to both the decoders.

\item \textbf{Encoding:} Encoder, upon observing $\svs^n$ at the output
of host source, sends messages $\rvw_1 \in \{1,2,\ldots,2^{nR_1}\}$ and
$\rvw_2 \in \{1,2,\ldots,2^{nR_2}\}$  by transmitting
the codeword $\rvx^n(\svs^n,\rvw_1,\rvw_2)$.
In this way, the codeword $\rvx^n$ is chosen and transmitted from
the encoder for a given host sequence $\rvs^n$,
and a given message pair $(\rvw_1,\rvw_2)$.

\item \textbf{Decoder~1:} Decoder 1, up on receiving
the channel output $\rvy^n$, looks for $\rvu^n(\svs^n,m_2)$ such that
$(\rvu^n(s^n,m_2),\rvy^n)\in T_{\epsilon}^n[\rvu,\rvy|\svs^n]$ for
all $\svs^n\in T_{\epsilon_1}^n[\rvs]$.
If a unique codeword $\rvu^n(\svs^n,m_2)$ exists,
Decoder~1 again looks for
$\rvx^n(\svs^n,m_1,m_2)$ such that
$(\rvx^n(\svs^n,m_1,m_2),\rvy^n)\in T_{\epsilon}^n[\rvx,\rvy|\svs^n,\rvu^n(\svs^n,m_2)]$.
If a unique codeword $\rvx^n(\svs^n,m_1,m_2)$ exists, Decoder~1
declares that $(\hat{\rvw}_1,\hat{\rvs}_2^n)=(m_1,\svs^n)$.
In this way, the message intended for Decoder~1 and
the host sequences are decoded at Decoder~1.

\item \textbf{Decoder~2:} Decoder 2, up on receiving
the channel output $\rvz^n$, looks for $\rvu^n(\svs^n,m_2)$ such that
$(\rvu^n(\svs^n,m_2),\rvz^n)\in T_{\epsilon}^n[\rvu,\rvz|\svs^n]$ for all
$\svs^n\in T_{\epsilon_1}^n[\rvs]$. If a unique codeword $\rvu^n(\svs^n,m_2)$
codeword exists, Decoder~2 declares that
$(\hat{\rvw}_2,\hat{\rvs}_1^n)=(m_2, \svs^n)$. Otherwise,
Decoder~2 declares an error. In this way, the message intended for
Decoder~2 and the host sequences are decoded at Decoder~2.

\item \textbf{Probability of error:} The average probability
of error is given by the following
{\allowdisplaybreaks
\begin{align}
P_e^n & = \sum_{(\svs^n)\in \calS^n }p(\svs^n)\mathrm{Pr}[\mathrm{error}|\svs^n] \nonumber \\
& \leq \sum_{\svs^n\not\in T_{\epsilon_1}^n[\rvs]}p(\svs^n)+\sum_{\svs^n\in
T_{\epsilon_1}^n[\rvs]}p(\svs^n)\mathrm{Pr}[\mathrm{error}|\svs^n],\nonumber \\
&= \sum_{\svs^n\not\in T_{\epsilon_1}^n[\rvs]}p(\svs^n)+\sum_{\svs^n\in
T_{\epsilon_1}^n[\rvs]}p(\svs^n)\mathrm{Pr}[\mathrm{E(1)\cup E((2)}|\svs^n],
\label{eqn:CaseC_pe_events}
\end{align}}
where $E(i)$ is the event that the error is made at Decoder~i, for $i=1,2.$
The first term, $\mathrm{Pr}[\svs^n\not\in T_{\epsilon_1}^n[\rvs]]$,
in the right hand side expression of (\ref{eqn:CaseC_pe_events})
goes to zero as $n \rightarrow \infty$ by Lemma~\ref{thm:ST_lemma1}.

Without loss of generality, it can be assumed that
the output of the host source is $\tilde{\svs}^n$, and $(\rvw_1,\rvw_2)=(1,1)$ is
being transmitted from the encoder.
Hence, the codeword $\rvx^n(\tilde{\svs}^n,1,1)$ is transmitted from the encoder.
Let $F_1$ be the event that $\tilde{\svs}^n \in T_{\epsilon_1}^n[\rvs]$ is output of 
the host source.

The following error events are considered to compute $\mathrm{Pr}[\mathrm{E(2)}|F]$
and can be made to approach zero as $n \rightarrow \infty$.
\begin{enumerate}
\item $E_1$: $(\rvu^n(\tilde{\svs}^n,1),\rvx^n(\tilde{\svs}^n,1,1),\rvy^n,\rvz^n) \not \in $
 $T_{\epsilon}^n[\rvs,\rvu,\rvx,\rvy,\rvz|\tilde{\svs}^n]$ under the event $F$.
 By using Lemma~\ref{thm:ST_lemma1}, we can show that $\mathrm{Pr}[E_{1}|F] \rightarrow 0$ as
 $n \rightarrow \infty$.
\item $E_2$: $(\rvu^n(\tilde{\svs}^n,m_2),\rvy^n) \in
T_{\epsilon}^n[\rvs,\rvu,\rvz|\tilde{\svs}^n]$ under the event $F \cap E_1^c$ for all $m_2 \neq 1$.
It can be shown that $\mathrm{Pr}(E_{2}|F) \rightarrow 0$ as $n \rightarrow \infty$
by using Lemma~\ref{thm:ST_lemma1} and  Lemma~\ref{thm:ST_lemma2}
if $0 \leq R_2 < \mathbb{I}(\rvu;\rvz|\rvs)$.
\item $E_3$: $(\rvu^n(\svs^n,m_2),\rvy^n) \in
T_{\epsilon}^n[\rvs,\rvu,\rvz|\svs^n]$ under the event $F \cap E_1^c$ for all $m_1$
and $\svs^n \neq \tilde{s}^n$.
It can be shown that $\mathrm{Pr}(E_{3}|F) \rightarrow 0$ as $n \rightarrow \infty$
by using Lemma~\ref{thm:ST_lemma1} and  Lemma~\ref{thm:ST_lemma2}
if $0 \leq R_2 < \mathbb{I}(\rvu,\rvs;\rvz)-\mathbb{H}(\rvs)$.
\end{enumerate}

From the all above error events, it can be concluded that
$\mathrm{Pr}[E(1)|F] \rightarrow 0$
as $n \rightarrow \infty$ if
$0 \leq R_2 < \mathbb{I}(\rvu,\rvs;\rvz)-\mathbb{H}(\rvs)$.
The following error events are considered to compute
$\mathrm{Pr}[\mathrm{E(1)}|F]$ and can be made to approach zero
as $n \rightarrow \infty$.
\begin{enumerate}
\item $E_4$:$(\rvu^n(\svs^n,m_2),\rvy^n) \in
T_{\epsilon}^n[\rvs,\rvu,\rvy|\svs^n]$  for $m_1 \neq 1$ or $\svs^n \neq \tilde{\svs}^n$.
By considering the error events similar to $E_2$ and $E_3$,
it can be shown that $\mathrm{Pr}(E_{4}|F,E_1^c) \rightarrow 0$ as
$n \rightarrow \infty$ if $0 \leq R_2 < \mathbb{I}(\rvu,\rvs;\rvy)-\mathbb{H}(\rvs)$.
\item $E_5$:$(\rvx^n(\tilde{\svs}^n,m_1,1),\rvy^n) \in
T_{\epsilon}^n[\rvs,\rvu,\rvx,\rvy|\tilde{\svs}^n,\rvu^n(\tilde{\svs}^n,1)]$
for $m_1 \neq 1$. It can be shown that $\mathrm{Pr}(E_{5}|F,E_1^c,E_4^c) \rightarrow 0$
as $n \rightarrow \infty$ if $0 \leq R_1 < \mathbb{I}(\rvx;\rvy|\rvs,\rvu)$.
\end{enumerate}

Then by using the union bound, $\mathrm{Pr}[\mathrm{E(1)\cup E(2)}|F]$ goes to zero as
$n \rightarrow \infty$ if rate pair $(R_1,R_2)$ satisfies
(\ref{eqn:capacity_schemeC}).
It can be concluded that $P_e^n \rightarrow 0$
as $n \rightarrow 0$ if rate pair
$(R_1,R_2)$ satisfies (\ref{eqn:capacity_schemeC}).
\item \textbf{Average distortions:} Since $(\rvx^n,\tilde{\svs}^n)$ is
jointly strongly typical with high probability and the distribution
belongs to $\calP(\Delta)$, it can be shown that the average
distortion $D^{(n)}$ associated with the generated code satisfies the
distortion constraint $\Delta$ as $n \rightarrow \infty$ as
in the Proof of Theorem~\ref{thm:innerbound_distributedRIE}.
\end{itemize}

\subsubsection{Converse}
We show that any sequence of
$(\lceil2^{nR_1}\rceil,\lceil2^{nR_2}\rceil,D^{(n)},n)$ codes,
i.e., $\rvx^n=f(\rvw_1,\rvw_2,\rvs^n)$, $g_{1,C'}^n(\rvy^n)=(\hat{\rvw}_1,\hat{\rvw}_2,\hat{\rvs}^n)$,
 and $g_{2,C'}^n(\rvz^n)=(\hat{\rvw}_2,\hat{\rvs}^n)$,
 with $\lim_{n \rightarrow \infty} P_e^n=0$ and $\lim_{n
\rightarrow \infty} D^{(n)} \leq \Delta$, the rate pair $(R_1,R_2)$
must satisfy (\ref{eqn:capacity_schemeC}) for some
$(\rvu,\rvs,\rvx,\rvy,\rvz) \in \calP(\Delta)$. Consider a
given code of block length $n$. The joint distribution on $\calW_1
\times \calW_2\times\calS^n \times \calX^n \times \calY^n \times
\calZ^n$ induced by the code is given by
\begin{eqnarray*}
 \lefteqn{p(\svw_1,\svw_2,\svs^n,\svx^n,\svy^n,\svz^n) =}\\
 & &\frac{1}{\lceil2^{nR_1}\rceil \lceil 2^{nR_2}\rceil}p(\svs^n)
 p(\svx^n|\svw_1,\svw_2,\svs^n) \\
 & &\times \prod_{i=1}^{n}p(\svy_j|\svx_{j},\svs_{j})p(\svz_j|\svy_j),
 \end{eqnarray*}
 where, $p(\svx^n|\svw_1,\svw_2,\svs^n)$ is $1$ if $\svx^n=
 f^n(\svw_1,\svw_2,\svs^n)$ and $0$  otherwise.

 We can bound the rate $R_1$ as follows:
 {\allowdisplaybreaks
 \begin{align}
 nR_1 \leq & \mathbb{H}(\rvw_1) \nonumber \\
      \stackrel{(a)}{=} & \mathbb{H}(\rvw_1|\rvw_2,\rvs^n) \nonumber \\
      = & \mathbb{H}(\rvw_1|\rvw_2,\rvs^n) -
      \mathbb{H}(\rvw_1|\rvw_2,\rvs^n,\rvy^n)
      + \mathbb{H}(\rvw_1|\rvw_2,\rvs^n,\rvy^n)\nonumber \\
      \stackrel{(b)}{\leq} &\mathbb{I}(\rvw_1 ; \rvy^n|\rvw_2,\rvs^n)
      + n \epsilon_n \nonumber \\
      = & \sum_{j=1}^n \mathbb{I}(\rvw_1 ; \rvy_j|\rvw_2,\rvs^n,\rvy^{j-1})
      + n \epsilon_n \nonumber \\
      = & \sum_{j=1}^n [\mathbb{H}(\rvy_j|\rvw_2,\rvs^n,\rvy^{j-1}) -
       \mathbb{H}(\rvy_j|\rvw_1,\rvw_2,\rvs^n,\rvy^{j-1})]
      + n \epsilon_n \nonumber \\
      \stackrel{(c)}{=} & \sum_{j=1}^n [\mathbb{H}(\rvy_j|\rvw_2,\rvs^n,\rvy^{j-1},\rvz^{j-1}) -
       \mathbb{H}(\rvy_j|\rvw_1,\rvw_2,\rvs^n,\rvy^{j-1}, \rvz^{j-1})]
      + n \epsilon_n \nonumber \\
      \stackrel{(d)}{\leq} & \sum_{j=1}^n [\mathbb{H}(\rvy_j|\rvw_2,\rvs^n,\rvz^{j-1}) -
       \mathbb{H}(\rvy_j|\rvw_1,\rvw_2,\rvs^n,\rvy^{j-1}, \rvz^{j-1},\rvx^n)]
      + n \epsilon_n \nonumber \\
      \stackrel{(e)}{=} & \sum_{j=1}^n [\mathbb{H}(\rvy_j|\rvw_2,\rvs^n,\rvz^{j-1}) -
       \mathbb{H}(\rvy_j|\rvx_j,\rvs_j)]
      + n \epsilon_n \nonumber \\
      \stackrel{(f)}{=} & \sum_{j=1}^n [\mathbb{H}(\rvy_j|\rvs_j,\tilde{\rvu}_j) -
       \mathbb{H}(\rvy_j|\rvx_j,\rvs_j)]
      + n \epsilon_n \nonumber \\
      = & \sum_{j=1}^n \mathbb{I}(\rvx_j ; \rvy_j|\rvs_j,\tilde{\rvu}_j)
      + n \epsilon_n, \label{eqn:temp1_capacity_schemeC}
      \end{align}}
\noindent
where, \\
(a) follows from the fact that $\rvw_1$, $\rvw_2$ and $\rvs^n$ are mutually independent, \\
(b) follows from Fano's inequality and $\epsilon_n \rightarrow 0$ as $n \rightarrow \infty$, \\
(c) follows from $\rvy_j \leftrightarrow (\rvw_2,\rvs^n, \rvy^{j-1}) \leftrightarrow \rvz^{j-1}$ and
$\rvy_j \leftrightarrow (\rvw_1,\rvw_2,\rvs^n, \rvy^{j-1}) \leftrightarrow \rvz^{j-1},$ \\
(d) follows from $\mathbb{H}(\rvy_j|\rvw_2,\rvs^n,\rvy^{j-1},\rvz^{j-1})
\leq \mathbb{H}(\rvy_j|\rvw_2,\rvs^n,\rvz^{j-1})$, and
$\rvx^n$ is a deterministic function of $(\rvw_1,\rvw_2,\rvs^n),$ \\
(e) follows from memoryless properties of the broadcast
channel, and \\
(f) follows from $\tilde{\rvu}_j :=\{\rvw_2,\rvs_{1}^{j-1}, \rvs_{j+1}^n\}.$

We can also bound the rate $R_2$ as follows:
 {\allowdisplaybreaks
 \begin{align}
 nR_2  \leq &\mathbb{H}(\rvw_2) \nonumber \\
       \stackrel{(a)}{\leq} & \mathbb{H}(\rvw_2,\rvs^n) -\mathbb{H}(\rvs^n) \nonumber \\
      \stackrel{(b)}{\leq} & \mathbb{I}(\rvw_2,\rvs^n;\rvz^n)-\mathbb{H}(\rvs^n) + n\epsilon_n \nonumber \\
      = & \sum_{j=1}^{n} [\mathbb{I}(\rvw_2,\rvs^n;\rvz_{j}|\rvz^{j-1})
      -\mathbb{H}(\rvs_j|\rvs^{j-1})] + n\epsilon_n \nonumber \\
      \stackrel{(c)}{=}& \sum_{j=1}^{n} [\mathbb{H}(\rvz^{j}|\rvz^{j-1})-
      \mathbb{H}(\rvz_{j}|\rvw_2,\rvs^n, \rvz^{j-1})- \mathbb{H}(\rvs_j)] +n\epsilon_n \nonumber \\
      \stackrel{(d)}{\leq}& \sum_{j=1}^{n} [\mathbb{H}(\rvz_{j})-
      \mathbb{H}(\rvz_{j}|\tilde{\rvu}_j,\rvs_j)- \mathbb{H}(\rvs_j)] +n\epsilon_n \nonumber \\
      = & \sum_{j=1}^{n} [\mathbb{I}(\tilde{\rvu}_j,\rvs_j;\rvz_j)- \mathbb{H}(\rvs_j)]+
      n\epsilon_n \nonumber \label{eqn:temp2_capacity_schemeC}
\end{align}}
\noindent
where, \\
(a) follows from the fact that $\rvw_1$, $\rvw_2$ and $\rvs^n$ are mutually independent, \\
(b) follows from Fano's inequality and $\epsilon_n \rightarrow 0$ as $n \rightarrow \infty$, \\
(c) follows from the fact that $\rvs^n$ is an i.i.d. random vector, \\
(d) follows from  $\mathbb{H}(\rvz_j|\rvz^{j-1})
\leq \mathbb{H}(\rvz_j)$, and $\tilde{\rvu}_j :=\{\rvw_2,\rvs_{1}^{j-1}, \rvs_{j+1}^n\}.$

We can then write
(\ref{eqn:temp1_capacity_schemeC}) and
(\ref{eqn:temp2_capacity_schemeC}) as
\begin{subequations} \label{eqn:temp3_capacity_schemeC}
\begin{align}
R_1 \leq & \mathbb{I}(\rvx;\rvy|\rvq,\rvs, \tilde{\rvu})+ \epsilon_n,  \\
R_2 \leq & \mathbb{I}(\tilde{\rvu},\rvs;\rvz|\rvq)
-\mathbb{H}(\rvs)+ \epsilon_n,
\end{align}
\end{subequations}
where $\rvq$ takes values in the set $\calQ \in \{1,2,\dots,n\}$ with
equal probability and the joint probability distribution on
$(\rvs,\rvq,\tilde{\rvu},\rvx,\rvy,\rvz)$ is
$p(\rvs=\svs,\rvq=\svq,\tilde{\rvu}=\tilde{\svu},\rvx=\svx)p(\svy|\svx,\svs)p(\svz|\svy),$
with
\begin{eqnarray*}
\lefteqn{p(\rvs=\svs,\rvq=\svq,\tilde{\rvu}=\tilde{\svu},\rvx=\svx)=}\\
& &p(\svs)p(\svq)p(\rvu_q=\tilde{\svu}|\svs,\svq)p(\rvx_q=\svx|\svs,\svq,\tilde{\svu}).
\end{eqnarray*}

Finally, we can write (\ref{eqn:temp3_capacity_schemeC}) as
\begin{align}
R_1 \leq & \mathbb{I}(\rvx;\rvy|\rvu,\rvs)+ n\epsilon_n, \nonumber \\
R_2 \leq & \mathbb{I}(\rvu,\rvs;\rvz)
-\mathbb{H}(\rvs)+ n\epsilon_n, \nonumber
\end{align}
where $\rvu:=(\rvq,\tilde{\rvu})$, since
$\mathbb{I}(\tilde{\rvu},\rvs;\rvz|\rvq) \leq \mathbb{I}(\rvq,\tilde{\rvu},\rvs;\rvz)$.

Given any $\delta >0$, the associated distortion $D^{(n)}$, for
sufficiently large $n$, satisfies
{\allowdisplaybreaks
\begin{align}
\Delta +\delta \geq & D^{(n)}\nonumber \\
                  =   & \mathbb{E}d(\rvx^n,\rvs^n) \nonumber \\
                  =   & \frac{1}{n} \sum_{j=1}^n \sum_{\svx,\svs}
		  p(\rvx_j=\svx,\rvs_j=\svs) d(\svx,\svs) \nonumber \\
		  =  & \sum_{\svx,\svs} p(\rvx=\svx,\rvs=\svs) d(\svx,\svs) \nonumber \\
		  = & \mathbb{E}d(\rvx,\rvs). \nonumber
\end{align}}
As $n \rightarrow \infty$ and $\delta \rightarrow 0$,
$(\rvu,\rvs,\rvx,\rvy,\rvz) \in \calP(\Delta)$ and $(R_1,R_2)
\in \calC_{C'}$.

\bibliographystyle{IEEEtran}
\bibliography{journal}

\end{document}

%% file: rv_defs.tex
\DeclareMathAlphabet{\eurm}{U}{eur}{m}{n}
\DeclareMathAlphabet{\mathbsf}{OT1}{cmss}{bx}{n}
\DeclareMathAlphabet{\mathssf}{OT1}{cmss}{m}{sl}
\DeclareMathAlphabet{\mathcsf}{OT1}{cmss}{sbc}{n}

\newcommand{\samplevalue}[1]{\eurm{\lowercase{#1}}}
\newcommand{\randomvalue}[1]{\eurm{\uppercase{#1}}}

\DeclareSymbolFont{bsfletters}{OT1}{cmss}{bx}{n}  
\DeclareSymbolFont{ssfletters}{OT1}{cmss}{m}{n}
\DeclareMathSymbol{\bsfGamma}{0}{bsfletters}{'000}
\DeclareMathSymbol{\ssfGamma}{0}{ssfletters}{'000}
\DeclareMathSymbol{\bsfDelta}{0}{bsfletters}{'001}
\DeclareMathSymbol{\ssfDelta}{0}{ssfletters}{'001}
\DeclareMathSymbol{\bsfTheta}{0}{bsfletters}{'002}
\DeclareMathSymbol{\ssfTheta}{0}{ssfletters}{'002}
\DeclareMathSymbol{\bsfLambda}{0}{bsfletters}{'003}
\DeclareMathSymbol{\ssfLambda}{0}{ssfletters}{'003}
\DeclareMathSymbol{\bsfXi}{0}{bsfletters}{'004}
\DeclareMathSymbol{\ssfXi}{0}{ssfletters}{'004}
\DeclareMathSymbol{\bsfPi}{0}{bsfletters}{'005}
\DeclareMathSymbol{\ssfPi}{0}{ssfletters}{'005}
\DeclareMathSymbol{\bsfSigma}{0}{bsfletters}{'006}
\DeclareMathSymbol{\ssfSigma}{0}{ssfletters}{'006}
\DeclareMathSymbol{\bsfUpsilon}{0}{bsfletters}{'007}
\DeclareMathSymbol{\ssfUpsilon}{0}{ssfletters}{'007}
\DeclareMathSymbol{\bsfPhi}{0}{bsfletters}{'010}
\DeclareMathSymbol{\ssfPhi}{0}{ssfletters}{'010}
\DeclareMathSymbol{\bsfPsi}{0}{bsfletters}{'011}
\DeclareMathSymbol{\ssfPsi}{0}{ssfletters}{'011}
\DeclareMathSymbol{\bsfOmega}{0}{bsfletters}{'012}
\DeclareMathSymbol{\ssfOmega}{0}{ssfletters}{'012}

\newcommand{\sva}{{\samplevalue{a}}}

\newcommand{\svb}{{\samplevalue{b}}}

\newcommand{\rvq}{{\randomvalue{q}}}	
\newcommand{\svq}{{\samplevalue{q}}}

\newcommand{\rvs}{{\randomvalue{s}}}	
\newcommand{\svs}{{\samplevalue{s}}}

\newcommand{\rvt}{{\randomvalue{t}}}	
\newcommand{\svt}{{\samplevalue{t}}}

\newcommand{\rvu}{{\randomvalue{u}}}	
\newcommand{\svu}{{\samplevalue{u}}}	

\newcommand{\rvv}{{\randomvalue{v}}}	
\newcommand{\svv}{{\samplevalue{v}}}	

\newcommand{\rvw}{{\randomvalue{w}}}	
\newcommand{\svw}{{\samplevalue{w}}}

\newcommand{\rvx}{{\randomvalue{x}}}	
\newcommand{\svx}{{\samplevalue{x}}}

\newcommand{\rvy}{{\randomvalue{y}}}	
\newcommand{\svy}{{\samplevalue{y}}}

\newcommand{\rvz}{{\randomvalue{z}}}	
\newcommand{\svz}{{\samplevalue{z}}}

\newcommand{\calC}{{\mathcal{C}}}

\newcommand{\calI}{{\mathcal{I}}}

\newcommand{\calP}{{\mathcal{P}}}
\newcommand{\calQ}{{\mathcal{Q}}}
\newcommand{\calR}{{\mathcal{R}}}
\newcommand{\calT}{{\mathcal{T}}}
\newcommand{\calS}{{\mathcal{S}}}
\newcommand{\calU}{{\mathcal{U}}}
\newcommand{\calV}{{\mathcal{V}}}
\newcommand{\calX}{{\mathcal{X}}}
\newcommand{\calY}{{\mathcal{Y}}}
\newcommand{\calW}{{\mathcal{W}}}
\newcommand{\calZ}{{\mathcal{Z}}}